%% file: main.tex
\pdfoutput=1

\documentclass[sigconf,screen = true, anonymous = false, nonacm]{acmart}

\AtBeginDocument{%
  \providecommand\BibTeX{{%
    \normalfont B\kern-0.5em{\scshape i\kern-0.25em b}\kern-0.8em\TeX}}}

\copyrightyear{2022}
\acmYear{2022}
\setcopyright{acmlicensed}
\acmConference[DeFi '22]{Proceedings of the 2022 ACM CCS Workshop on Decentralized Finance and Security}{November 11, 2022}{Los Angeles, CA, USA}
\acmBooktitle{Proceedings of the 2022 ACM CCS Workshop on Decentralized Finance and Security (DeFi '22), November 11, 2022, Los Angeles, CA, USA}
\acmPrice{15.00}
\acmDOI{10.1145/3560832.3563435}
\acmISBN{978-1-4503-9882-4/22/11}

\usepackage{subcaption}
\usepackage{tikz}
\usepackage{lipsum}
\usepackage[export]{adjustbox}
\usepackage{textcomp}
\begin{document}

\title{Exploring Price Accuracy on Uniswap V3 in Times of Distress}

\author{Lioba Heimbach}
\affiliation{%
\institution{ETH Zürich}
  \country{Switzerland}}
\email{hlioba@ethz.ch}
\author{Eric Schertenleib}
\affiliation{%
\institution{ETH Zürich}
  \country{Switzerland}}
\email{ericsch@ethz.ch}

\author{Roger Wattenhofer}
\affiliation{%
\institution{ETH Zürich}
  \country{Switzerland}}
\email{wattenhofer@ethz.ch}
 

\begin{abstract}
Financial markets have evolved over centuries, and exchanges have converged to rely on the order book mechanism for market making. Latency on the blockchain, however, has prevented \emph{decentralized exchanges (DEXes)} from utilizing the order book mechanism and instead gave rise to the development of market designs that are better suited to a blockchain. Although the first widely popularized DEX, Uniswap V2, stood out through its astonishing simplicity, a recent design overhaul introduced with Uniswap V3 has introduced increasing levels of complexity aiming to increase capital efficiency.  

In this work, we empirically study the ability of Unsiwap V3 to handle unexpected price shocks. Our analysis finds that the prices on Uniswap V3 were inaccurate during the recent abrupt price drops of two stablecoins: UST and USDT. We identify the lack of agility required of Unsiwap V3 liquidity providers as the root cause of these worrying price inaccuracies. Additionally, we outline that there are too few incentives for liquidity providers to enter liquidity pools, given the elevated volatility in such market conditions. 
\end{abstract}




\maketitle

\section{Introduction}
 
While the first stock markets date back at least to the 17th century and have thus a long track record, their novel \emph{decentralized finance (DeFi)} counterpart: \emph{decentralized exchanges (DEXes)} are still in their teething phase. Most financial markets rely on an order book mechanism, where constant buy and sell orders determine the price in an auction-like fashion. The price lies between the highest \emph{bid} (buy order) and lowest \emph{ask} (sell order) and can adapt rapidly. 

DEXes are built by smart contracts hosted on the blockchain, mainly Ethereum~\cite{wood2014ethereum}, and the significant latency between subsequent blocks has generally prohibited the adoption of the order book mechanism on the blockchain. Instead DEXes, generally function as \emph{constant function market makers (CFMMs)}. CFMMs rely on a predetermined price curve for on-chain market making. Individual liquidity providers aggregate liquidity in what is referred to as a liquidity pool for each tradeable cryptocurrency pair. Liquidity providers are renumerated for their service through transaction fees but bear the risk of high volatility. 

The first widely popular CFMM was Uniswap V2~\cite{adams2020uniswap}. Uniswap V2 functions as a \emph{constant product market maker (CPMM)}, where liquidity providers supply liquidity on the entire price range. The CPMM ensures that the product of the pool's two reserved cryptocurrencies remains constant. Thus, executed trades directly cause price shifts in the market. Large price differences between markets give rise to arbitrage opportunities which arbitrage traders are expected to exploit until the arbitrage opportunity vanishes.  

Curve~\cite{2021curve} later proposed a CFMM specialized in stablecoin (cryptocurrencies designed to have stable prices) liquidity pools. Stablecoins are often pegged to the US\$ and should thus trade at an equal price. Curve's CFMM adapts the CPMM to reduce the price change for a given trade size as long as stablecoins are trading near their equal price. In a recent design overhaul, Uniswap implemented an updated protocol V3~\cite{adams2021uniswap}, where liquidity providers have great freedom in selecting the price range for which they wish to facilitate trading. This update was marketed, in part,  to be specially tailored for stablecoin pairs, as the more concentrated liquidity distribution at the intended price should increase capital efficiency.

For stablecoin pairs on Uniswap V3, the liquidity is highly concentrated around the equal price point of the two assets in the pool, i.e., both assets trade at 1 US\$. Thus, as soon as the price moves outside an extremely narrow price range and the liquidity is not adjusted accordingly, there is little to no liquidity left to support trading. The concentrated liquidity CFMM introduced by Uniswap V3 was tested in early May 2022. First, UST, a stablecoin with a market capitalization of 18.7B US\$~\cite{2022terrausd}, lost its peg of 1 US\$ on 7 May and plummeted to a couple of cents within days. USDT, the largest stablecoin with a market capitalization of 83.2B US\$, also lost its peg of 1 US\$ and temporarily traded at 94~cents on 11 May. While USDT largely recovered from its price drop, UST did not. 

In this work, we empirically analyze the ability of the concentrated liquidity CFMM utilized by Uniswap V3 to handle such unexpected price shocks and find that Uniswap V3 markets were not able to adjust to the dramatic price drops of UST and USDT in early May 2022. We further conclude that the lack of sophistication and agility of Uniswap V3 liquidity providers is largely responsible for the market's price inaccuracies. Additionally, we find that once the supposedly stable price of a pair turns volatile, there is a lack of incentives for liquidity providers to create positions that would facilitate trading at the current price, given the significant risk. 

\section{Related Work}

Limit order books dominate market making in traditional markets. Market making is largely in the hands of professional market makers, i.e., specialized firms. The professionals are tasked to ensure the accuracy of the price, and their ability to do so determines their own financial success. The high latency and transaction cost on the blockchain explain why CFMMs, thanks to their simplicity, have generally been used for market making. Yet, the recent introduction of concentrated liquidity CFMMs has brought market making on the blockchain closer to traditional market making as liquidity providers must select a price range. We study the price accuracy of concentrated liquidity CFMMs when prices swing unexpectedly.

Angeris et al.~\cite{angeris2019analysis} were the first to formally study CPMMs and show that these markets must closely follow the price of the reference market. In subsequent work, Angeris et al.~\cite{angeris2020improved} extend the analysis to more general CFMMs. In contrast to their work, we empirically investigate the ability of Uniswap V3, an updated CFMM protocol, to track the price of the reference market. 

Chitra et al.~\cite{chitra2021liveness} compare on-chain CFMMs to conventional limit order books during periods of liveness loss. They find that both order books and concentrated liquidity CFMMs perform worse than unbounded CFMMs under loss of liveness. As opposed to studying concentrated liquidity CFMMs under liveness loss, we analyze their ability to track the price of the primary market during unexpected stablecoin price shocks. 

Berg et al.~\cite{Berg2022empirical} study the presence of market inefficiencies on Uniswap V2 by searching for and identifying cyclic arbitrage opportunities; indicative of price inaccuracies in the market. In our work, we measure price inaccuracies of Uniswap V3 by comparing the market's price to that of the primary market. Our analysis focuses on a time characterized by stablecoin price shocks.

Recently, several studies have emerged~\cite{neuder2021strategic,Fritsch2021concentrated,loesch2021impermanent,Heimbach2022risks} that focus on the novel CFMM design pioneered by Uniswap V3. These studies generally focus on performance and strategies for Uniswap V3 liquidity providers. On the other hand, we present the first empirical study on the ability of markets utilizing the concentrated liquidity model to track the price of the primary market. 

Multiple studies of cryptocurrency liquidity shocks in traditional finance have been published~\cite{tang2022liquidity,corbet2022cryptocurrency,griffin2020bitcoin,kliber2018price}. Conversely, our analysis focuses on liquidity shocks on concentrated liquidity CPMMs and the role their unique market design plays therein. 

\section{Constant Function Market Maker}
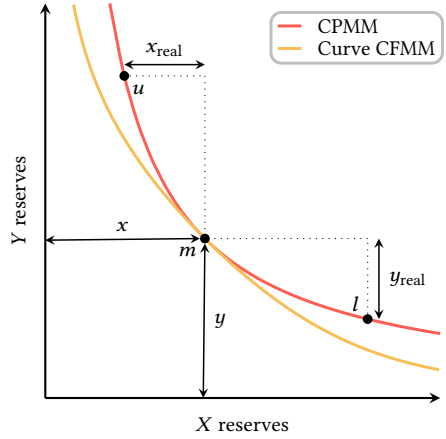
\begin{figure}[tbp]
  \centering
    \input{TikzFiles/virtual.tikz}\vspace{-14pt}
    
    \caption{We draw the CPMM (red line) and Curve CFMM price curve (purple line). The CPMM price curve ensures the product invariant, and the Curve CFMM guarantees the Curve invariant. The (virtual) reserves of $X$- and $Y$-token at point $m$ are $x$ and $y$, while the marginal is given by the gradient of the slope. Note that Curve's price curve is less convex and, thus, has lower price slippage near the equal price. In turn, the price curve is steeper away from $m$.} \label{fig:curve}\vspace{-9pt}
 \end{figure}%
 
In CFMMS, trades execute automatically, and a predefined invariant controls the cryptocurrency prices. CFMMs aggregate the liquidity from many individual liquidity providers in \emph{liquidity pools}, that are, smart contracts on the blockchain. Liquidity pools are created for each tradeable group of cryptocurrencies. Some DEXes restrict the number of cryptocurrencies in a liquidity pool to two, while others allow for liquidity pools with varying numbers of tokens.

\subsection{Constant Product Market Maker}
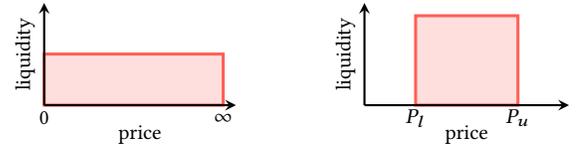
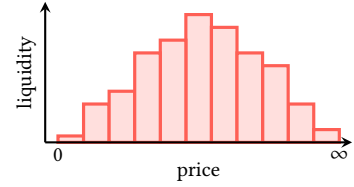
\begin{figure}[t]
\centering
  
  \begin{subfigure}{0.49\linewidth}
  
  \centering
   \input{TikzFiles/liquidityv2.tikz}\vspace{-8pt}
    \caption{Uniswap V2 liquidity position} \label{fig:liquidityv2}
  \end{subfigure}%
  \hfill
  \begin{subfigure}{0.49\linewidth}
  \centering
    \input{TikzFiles/liquidityv3.tikz}\vspace{-8pt}
    \caption{Uniswap V3 liquidity position} \label{fig:liquidityv3}
\end{subfigure}\vspace{3pt}

\begin{subfigure}{\linewidth}
  \centering
    \input{TikzFiles/liquiditys.tikz}\vspace{-8pt}
    \caption{collection of Uniswap V3 liquidity positions} \label{fig:liquiditys}\vspace{-10pt}
\end{subfigure}
\caption{Visualization of liquidity allocation in (concentrated liquidity) CPMMs. While a liquidity position's liquidity is distributed uniformly on Uniswap V2 (cf. Figure~\ref{fig:liquidityv2}), each Uniswap V3 liquidity position specifies a price interval $[P_l,P_u]$ to distribute liquidity across (cf. Figure~\ref{fig:liquidityv3}). The liquidity distribution of a collection of positions across the price range on Uniswap V3 is non-uniform (cf. Figure~\ref{fig:liquiditys}).} \label{fig:liquidity}\vspace{-10pt}
\end{figure}

The most widely adopted subclass of the CFMM is the CPMM. Uniswap V3, its predecessors Uniswap V1 and V2~\cite{adams2020uniswap} as well as SushiSwap~\cite{2021sushiswap} functions as a CPMM. A CPMM ensures the product between the pool's reserves remains constant during trading. The state of a liquidity pool with two cryptocurrencies, thus, moves along the red price curve shown in Figure~\ref{fig:curve}. 

Consider a liquidity pool $X-Y$ for $X$-tokens and $Y$-tokens that holds $x$ $X$-tokens and $y$ $Y$-tokens. The pool's \emph{marginal price} is then given by $P=y/x$ and the pool's \emph{liquidity} is $L=\sqrt{x\cdot y}$~\cite{adams2020uniswap}. Note that in the original CPMM design, liquidity is unbounded, i.e., it supports trading on the pool's entire price range $(0,\infty)$ (cf. Figure~\ref{fig:liquidityv2}).

Adams et al.~\cite{adams2021uniswap} pioneered an overhauled CPMM design with Uniswap V3: a concentrated liquidity CPMM. A liquidity provider concentrates their liquidity in their chosen price range $[P_l, P_u]$, see Figure~\ref{fig:liquidityv3} for a sample liquidity position. 
Importantly, each liquidity position only facilitates trading in the specified price range $[P_l, P_u]$ and the liquidity provider only collects fees if trading occurs within this range. Therefore, liquidity providers intend to select their range around a price they view as a realistic future market price. The aggregation of the set of individual liquidity positions of the pool then gives the liquidity distribution of the pool across the entire price range (cf. Figure~\ref{fig:liquiditys}).

To describe the pool's behavior on a price interval $[P_l, P_u]$ with constant liquidity, Uniswap V3 utilizes the concept of \emph{virtual reserves}. Virtual reserves simulate trading on the price interval as if the liquidity distribution on the entire price range $(0,\infty)$ is constant (cf. Figure~\ref{fig:liquidityv2}) and matches that of the interval $[P_l,P_u]$. The virtual reserves behave according to the constant product price curve, as shown in Figure~\ref{fig:curve}. Thus, the protocol ensures that the product of the virtual reserves $x$ and $y$ stays constant, i.e., $x\cdot y=L^2$. Here, $L$ is the liquidity reserved on the price interval $[P_l,P_u]$ and the pool's marginal price is given by $P= y/x$~\cite{adams2021uniswap}. The following relationship then holds between virtual reserves, real reserves, and liquidity: $$ x_\text{real} =  x- \frac{L}{\sqrt{P_u}} =  \frac{L}{\sqrt{P}} - \frac{L}{\sqrt{P_u}} \qquad y_{\text{real}}=y-L \sqrt{P_l} =L \sqrt{P}-L \sqrt{P_l}.$$ 
We emphasize that the liquidity on the price interval $[P_l, P_u]$ must only support trading within its price boundaries. The real reserves for the $X$ token are fully depleted at the upper price limit $P_u$ (cf. Figure~\ref{fig:liquidityv3}). The same holds for $Y$-tokens at the lower price boundary. 

Note that most CPMMs charge a percentage fee $f$ on the trade input. While all Uniswap V2 pools have the same percentage fee ($0.3\%$), each Uniswap V3 pool has a predefined percentage fee from the following set: $\{0.01\%, 0.05\%,0.3\%,1\%\}$. 

\subsection{Curve Constant Function Market Maker}
Curve is a DEX that uses a unique CFMM that is supposed to be better suited for stablecoin pairs as the price curve is \emph{flatter} near the peg but steeper further away compared to Uniswap V2~\cite{egorov2019stableswap} (cf. Figure~\ref{fig:curve}). However, as with Uniswap V2, liquidity is distributed across the entire price range, so we expect a qualitatively similar behavior in the event of price turbulences.

\section{Data Collection}
We analyze price data between various stablecoin pairs between 7 May and 12 May 2022. In particular, our analysis focuses on two stablecoins: UST and USDT. The price of the UST began to falter on 7 May and fell to 30 cents by 11 May. UST never fully recovered from the price drop and currently trades at a couple of cents. On 12 May, the price of USDT, the biggest stablecoin in terms of market capitalization, also dramatically dropped and landed 6\% below its 1 US\$ peg. The price, however, recovered, and USDT has since been trading only slightly below its peg.   

To analyze the impact of these unexpected price drops on the concentrated liquidity model introduced by Uniswap V3, we compare the price curve of stablecoin pairs, including UST or USDT, to that of Uniswap V2, Curve, and Binance. We analyze all stablecoin pairs between UST, USDT, and USDC with significant liquidity. USDC's price was stable in mid-May and is, therefore, used as a price reference for the other two stablecoins.

On Uniswap V3 we consider the following pools: USDT-USDC ($f = \{0.01\%,0.05\%\}$), UST-USDC ($f = \{0.01\%,0.05\% \}$), and UST-USDT ($f = 0.05\%$). These include all pools between the three analyzed cryptocurrencies that had more than 30'000 US\$ liquidity on 6 May 2022. We include a pair's price from Uniswap V2 if a pool with more than 30'000 US\$ liquidity exists. Curve price data is inferred from the \emph{3pool} liquidity pool, the biggest stablecoin pool on Curve that trades USDC, USDT, and DAI, as well as the \emph{ust} liquidity pool, which trades UST, USDC, USDT, and DAI. For Binance, we collect the price data for UST-USDT and USDC-USDT, from which we infer UST-USDC price data. Binance is by far the leading cryptocurrency market in terms of volume~\cite{2022exchangeranking}, and thus we will consider its price the most reliable for the cryptocurrencies that are traded thereon. 

We launch an Erigon~\cite{2022erigon} client to collect data from the Ethereum blockchain between block 14'726'564 (first block of 7 May) and block 14'764'083 (last block of 12 May). In particular, we filter event logs for all events related to Uniswap V2 and V3 as well as Curve. Price data from Binance is obtained from their market data API~\cite{2022bianancemarketdate} and downloading trade data for the relevant trading pairs.

\section{Analysis}
\begin{figure*}[t]
\centering

    \begin{subfigure}{0.48\linewidth}
  
        \centering
        \includegraphics[scale =1]{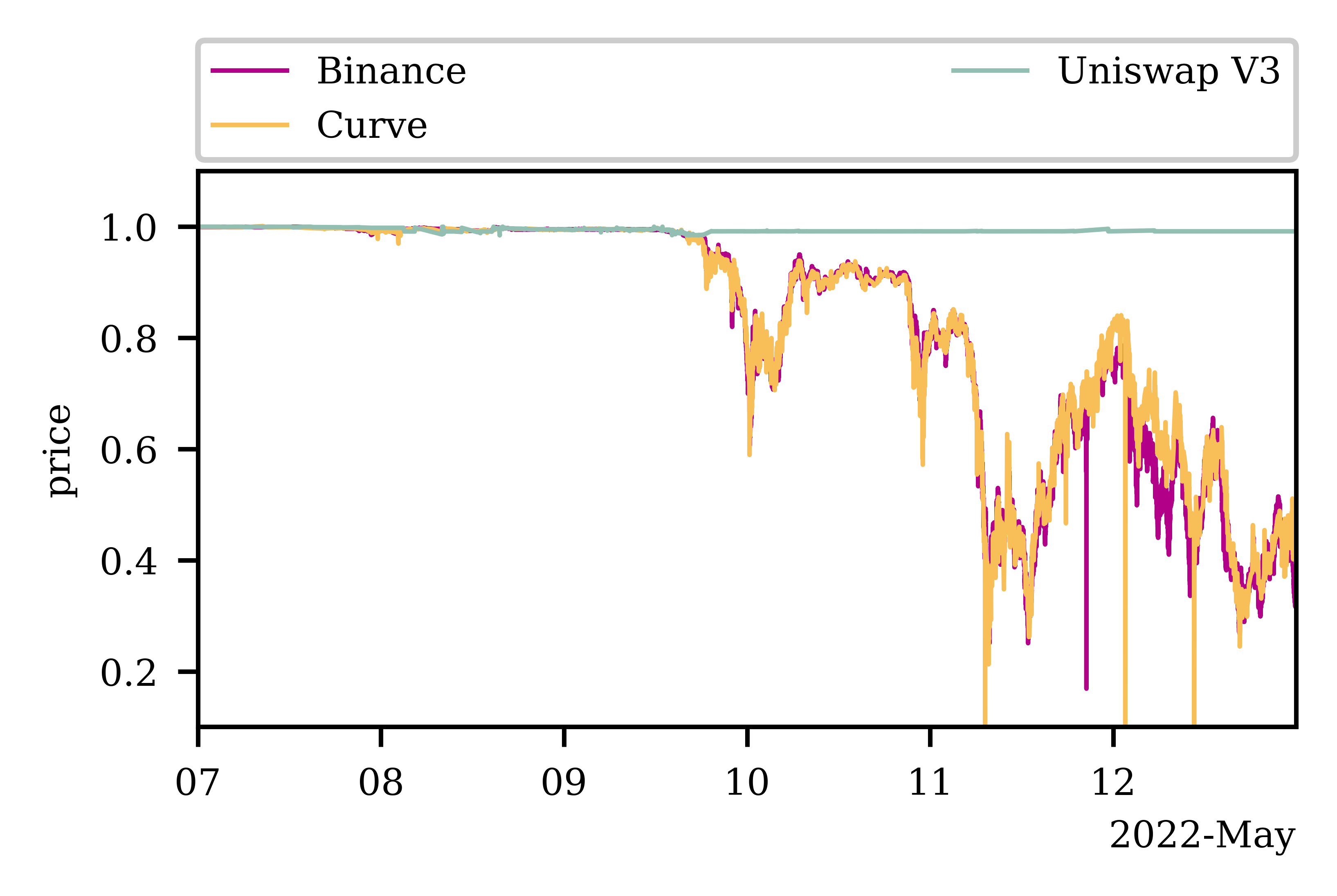}\vspace{-13pt}
        \caption{UST-USDC (f = 0.01\%)} \label{fig:priceaccUSDC-UST-0.0001}
    \end{subfigure}%
    \hfill
    \begin{subfigure}{0.48\linewidth}
        \centering
        \includegraphics[scale =1]{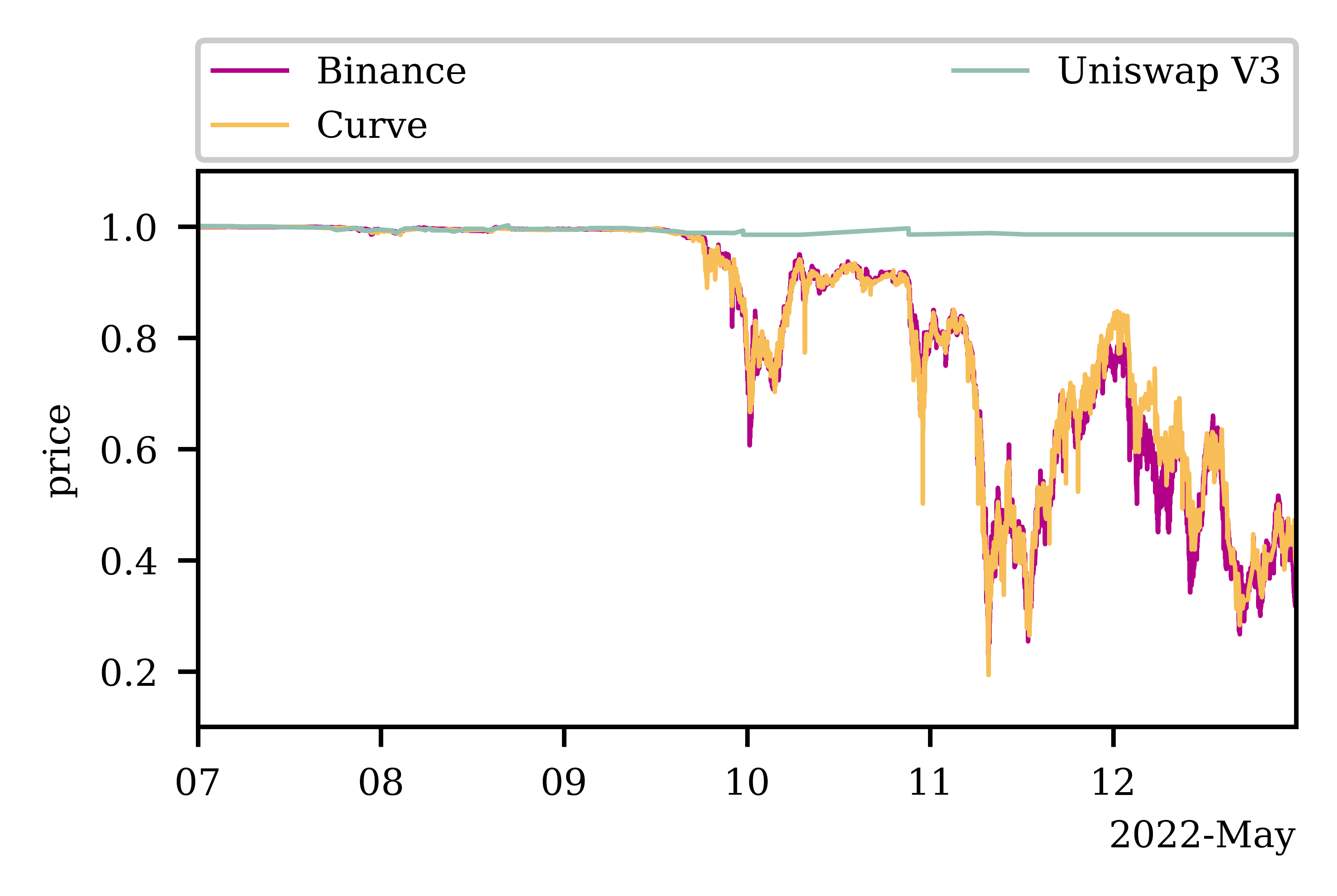}\vspace{-13pt}
        \caption{UST-USDT (f = 0.05\%)} \label{fig:priceaccUST-USDT-0.0005}
    \end{subfigure}\vspace{-3pt}
  
    \begin{subfigure}{0.48\linewidth}
      
        \centering
        \includegraphics[scale =1]{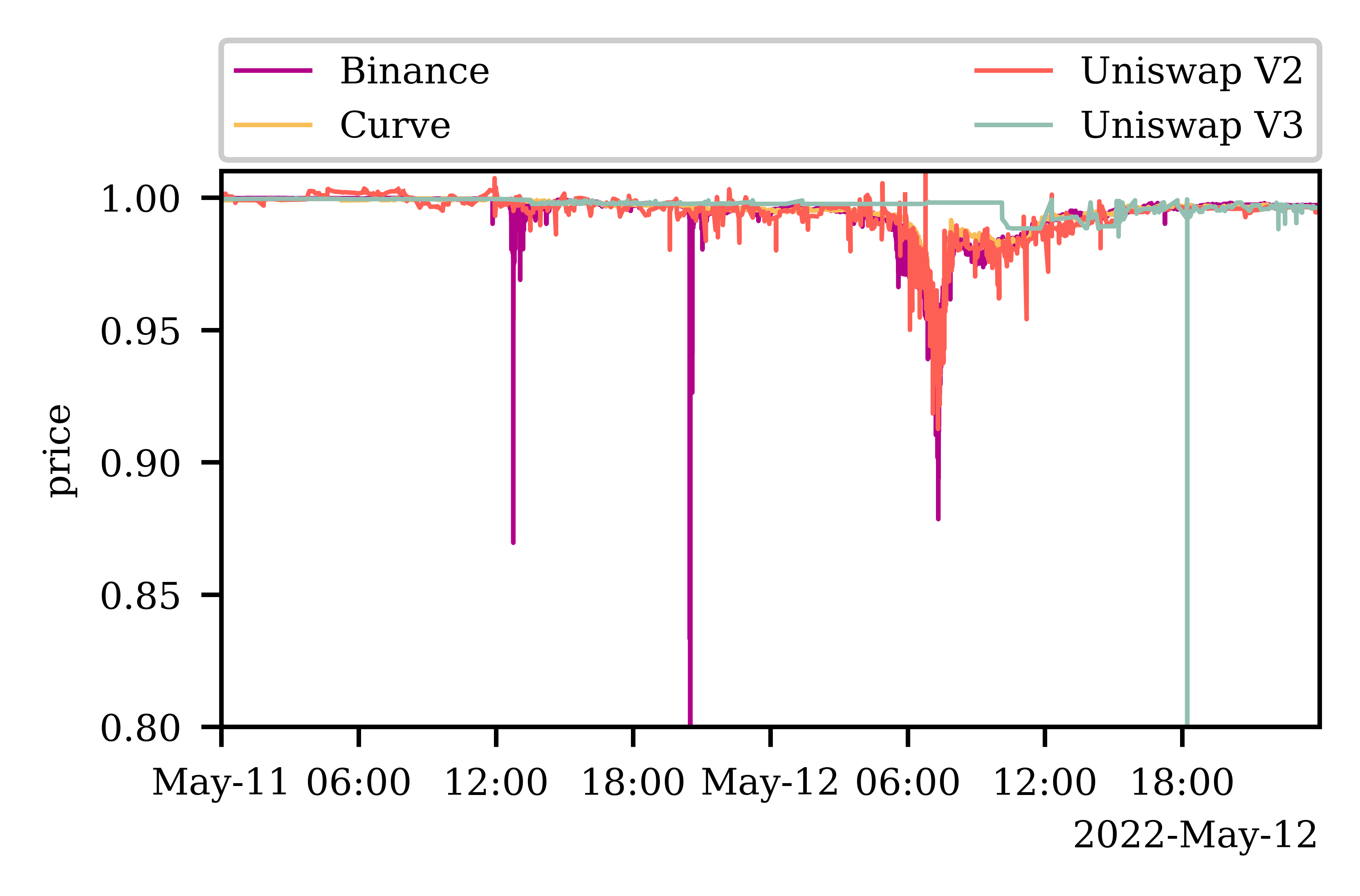}\vspace{-13pt}
        \caption{USDT-USDC (f = 0.01\%)} \label{fig:priceaccUSDC-USDT-0.0001}
    \end{subfigure}%
    \hfill
     \begin{subfigure}{0.48\linewidth}
        \centering
        \includegraphics[scale =1]{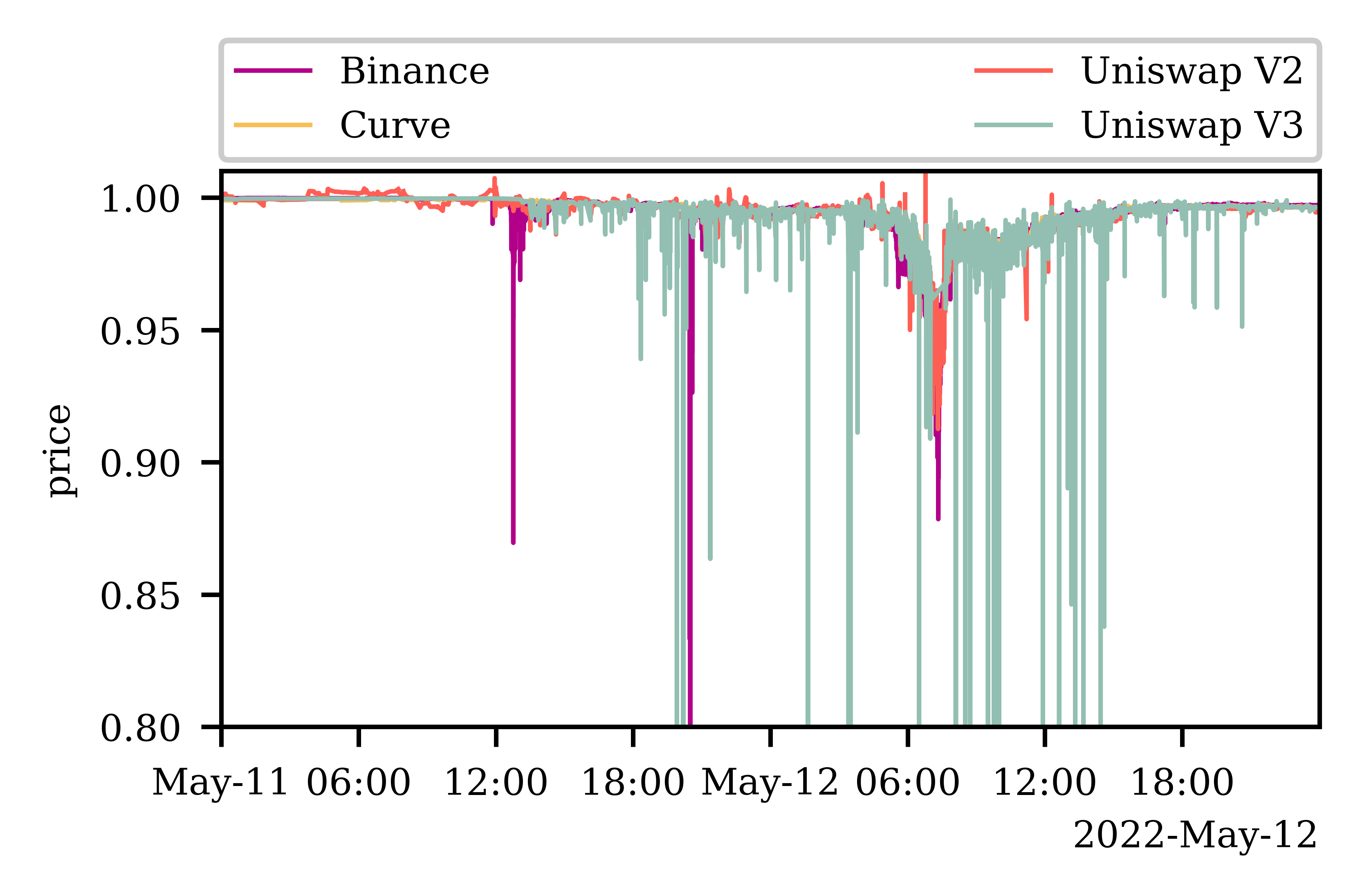}\vspace{-13pt}
        \caption{USDT-USDC (f = 0.05\%)} \label{fig:priceaccUSDC-USDT-0.0005}
    \end{subfigure}\vspace{-8pt}
\caption{We plot the price between various stablecoin pairs on Uniswap V3, Uniswap V2 (where applicable), Curve, and Binance. Figures~\ref{fig:priceaccUSDC-UST-0.0001} and~\ref{fig:priceaccUST-USDT-0.0005} show the price of UST in terms of USDC and UST, respectively, in the Uniswap V3 pools along with the price curves from Curve and Binance. In Figures~\ref{fig:priceaccUSDC-USDT-0.0001} and~\ref{fig:priceaccUSDC-USDT-0.0005} we plot the price of USDT in terms of USDC for the Uniswap V3 pool with fees 0.01\% and 0.05\% respectively. The corresponding price curves for Uniswap V2, Curve, and Binance are included.} \label{fig:priceacc}\vspace{-10pt}
\end{figure*}

We begin the analysis by studying the price of various stablecoin pools on Uniswap V3 and comparing them with the price on Binance, which we consider to be the most accurate, as well as Uniswap V2 and Curve. In Figure~\ref{fig:priceacc}, we plot these price curves for four of the five Uniswap V3 pools we analyze. Note that we only plot the price curve for the UST-USDC pool with the smaller fee tier as the behavior of the other UST-USDC pool is very similar. 

We immediately notice that the price between the pool's two assets is less likely to move for the analyzed Uniswap V3 pairs than the price on Uniswap V2, Curve, or Binance. For instance, we plot the price of UST in terms of USDC in Figure~\ref{fig:priceaccUSDC-UST-0.0001} and USDT in Figure~\ref{fig:priceaccUST-USDT-0.0005}. We observe that UST's price on Uniswap V3 never falls significantly below its peg of 1 US\$. On Binance, however, the price of UST dropped from its peg to around 0.30 cents in the same time frame. The price on Curve (cf. Figure~\ref{fig:priceaccUST-USDT-0.0005}) tracks the price on Binance closely, as opposed to Uniswap V3's price. It appears that there was no significant liquidity available for the Uniswap V3 stablecoin pools that held UST as one of their assets to continue to facilitate trading. Considering the very significant and unexpected price drop of UST, it would not be uncommon for financial markets to eventually stop trading assets given such events. Binance, for instance, stopped trading UST on 13 May due to its dramatic price drop~\cite{2022LUNA}. However, looking at Uniswap V3, we find that it could not even follow the initial smaller price fluctuations on 9 May.

The picture becomes even more startling when we study the price curves of the two Uniswap V3 pools trading USDT and USDC (cf. Figures~\ref{fig:priceaccUSDC-USDT-0.0001} and~\ref{fig:priceaccUSDC-USDT-0.0005}). USDT experienced an abrupt but less significant price drop than UST on 11 and 12 May, from which it has since largely recovered. When analyzing Figures~\ref{fig:priceaccUSDC-USDT-0.0001} and~\ref{fig:priceaccUSDC-USDT-0.0005}, we observe that while Uniswap V3 struggles significantly to follow these small but unexpected price fluctuations, Curve and Uniswap V2 can easily keep up. In the USDT-USDC ($f=0.01\%$) pool, the price of Uniswap V3 appears to react only after 36 hours. By then, the price of USDT had already recovered significantly. However, the price of USDT in the Uniswap V3 pool still experienced a sudden and short price drop on 12 May at 18:00. Two similar price drops occurred on Binance a day earlier and in all three cases could be caused by low liquidity levels. The USDT-USDC ($f=0.05\%$) pool reacted much more quickly to the USDT price drop. However, the liquidity pool appears to overreact, i.e., experience sudden price drops repeatedly. The USDT price drops significantly below the price observed in the other three markets a couple of dozen times.

\begin{figure*}[t]
\centering
    \begin{subfigure}{0.48\linewidth}
  
        \includegraphics[scale =1,left]{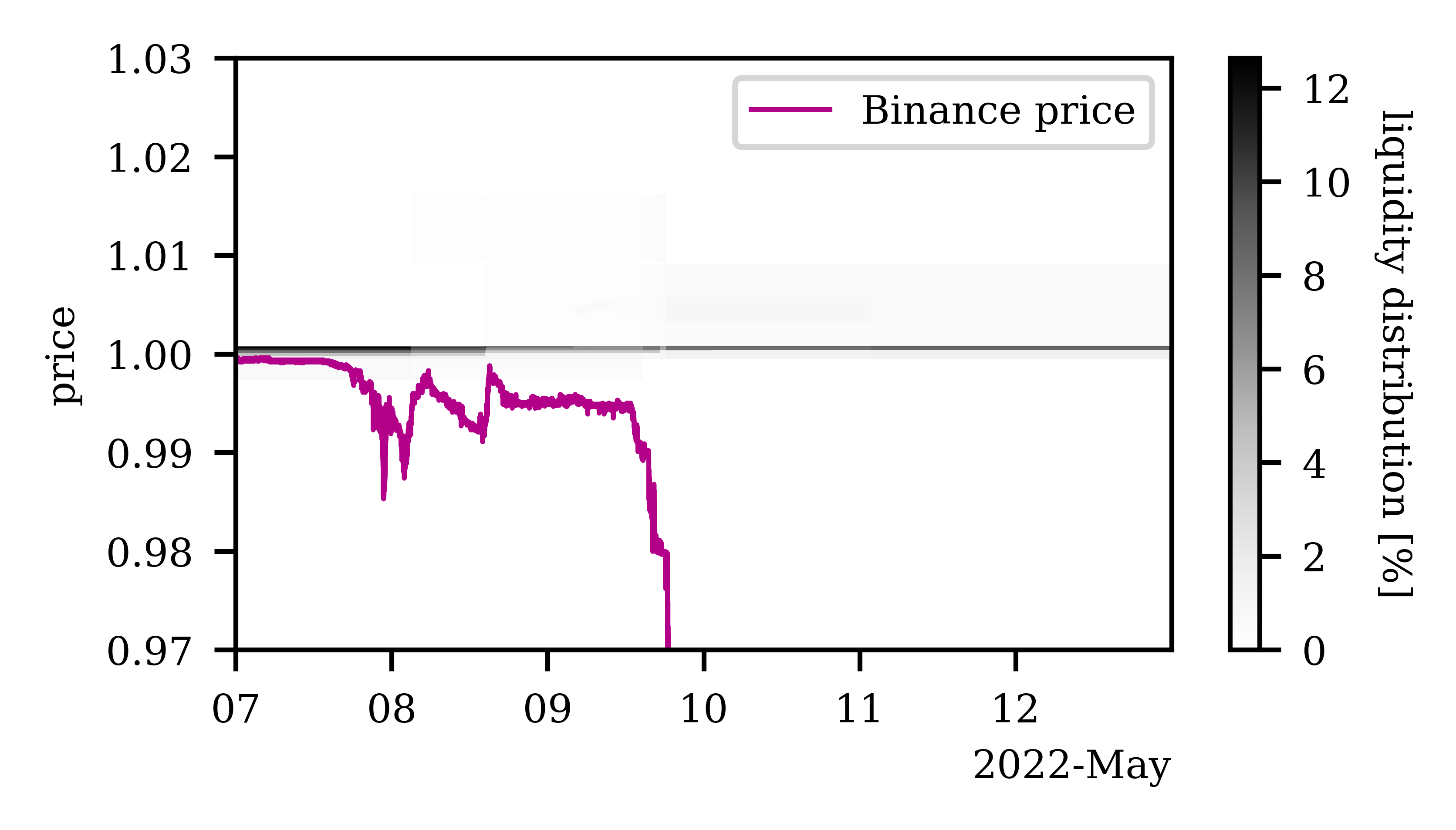}\vspace{-13pt}
        \caption{UST-USDC (f = 0.01\%) }
        \label{fig:priceandliquidityUSDC-UST-0.0001}
    \end{subfigure}%
    \hfill
    \begin{subfigure}{0.48\linewidth}
  
        \includegraphics[scale =1,left]{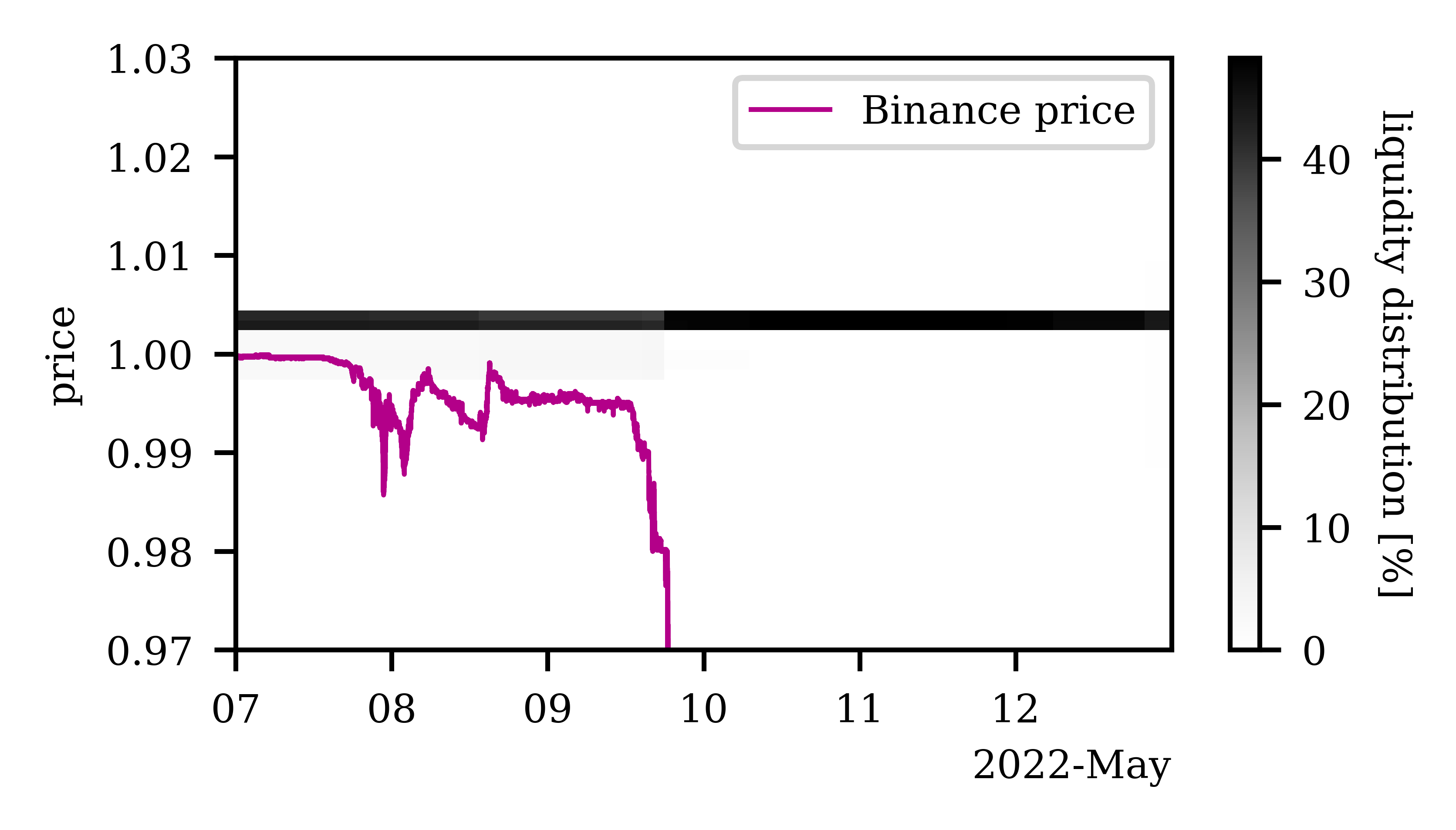}\vspace{-13pt}
        \caption{UST-USDT (f = 0.05\%)} 
        \label{fig:priceandliquidityUST-USDT-0.0005}
    \end{subfigure}\vspace{-3pt}
    
    \begin{subfigure}{0.48\linewidth}
  
        \includegraphics[scale =1,left]{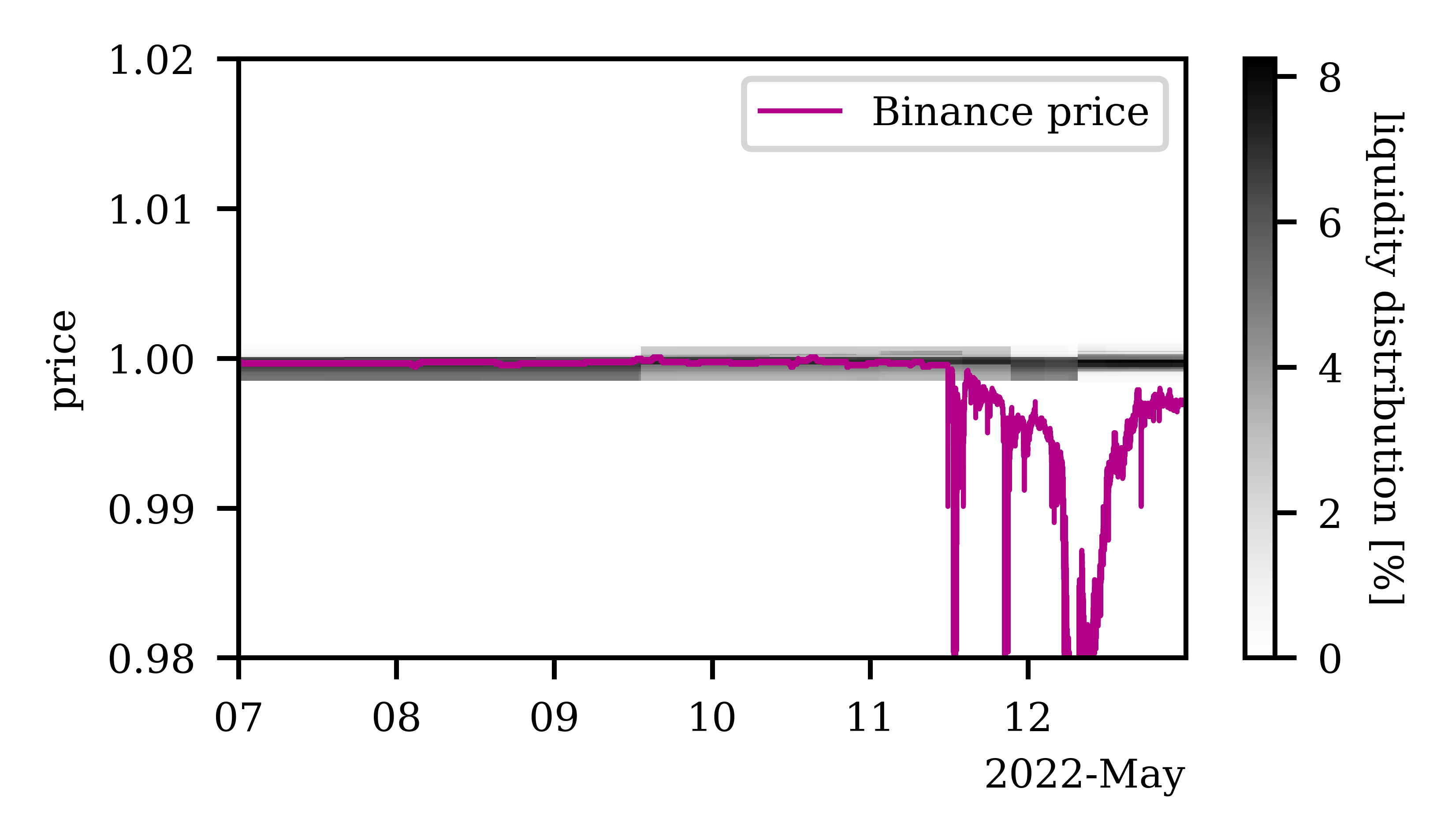}\vspace{-13pt}
        \caption{USDT-USDC (f = 0.01\%)} 
        \label{fig:priceandliquidityUSDC-USDT-0.0001}
    \end{subfigure}%
    \hfill
    \begin{subfigure}{0.48\linewidth}
  
        \includegraphics[scale =1,left]{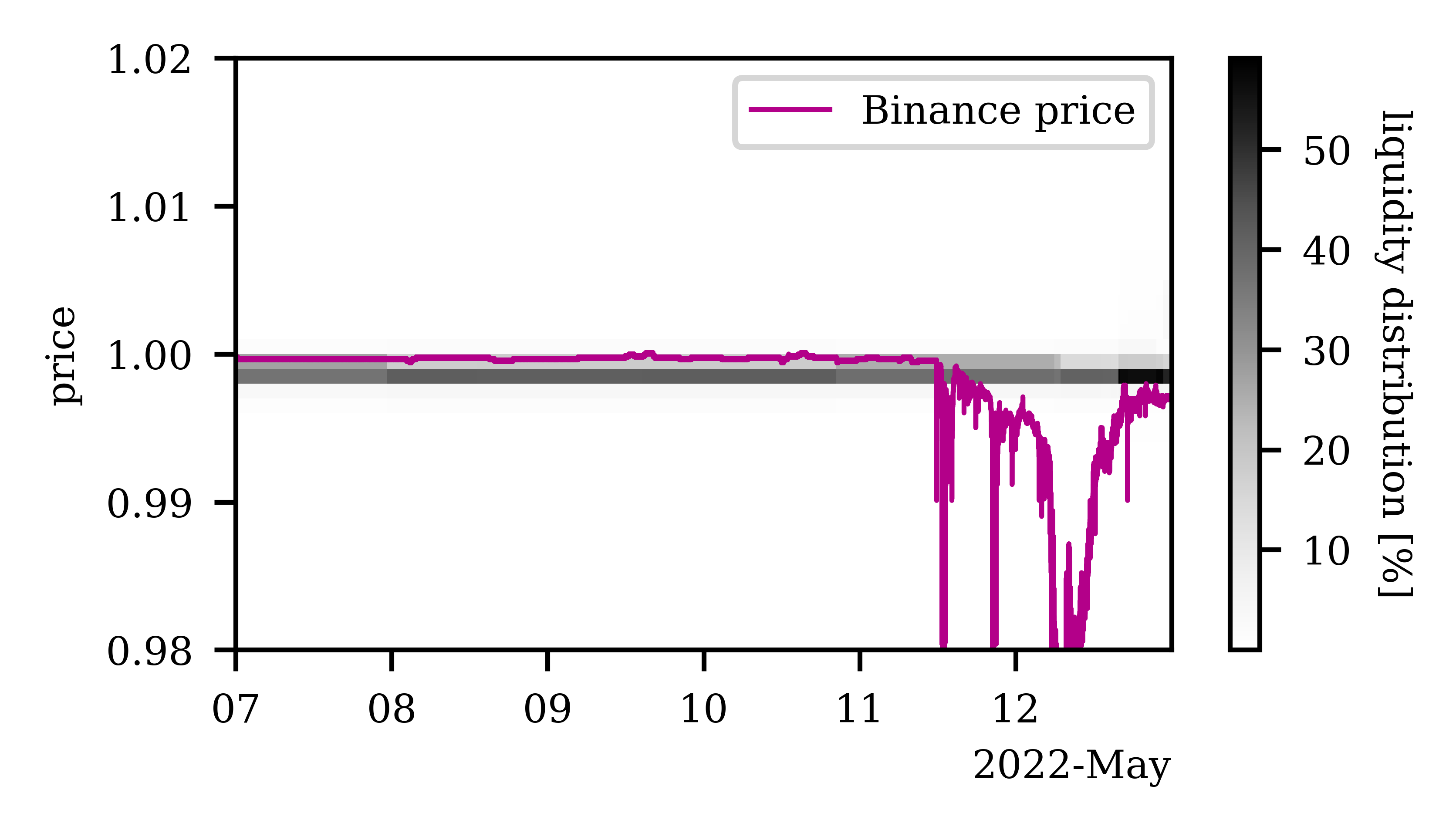}\vspace{-13pt}
        \caption{USDT-USDC (f = 0.05\%)}
        \label{fig:priceandliquidityUSDC-USDT-0.0005}
    \end{subfigure}
    \vspace{-8pt}
\caption{We visualize the liquidity distribution on the price interval over time and include the price curve of the primary market (Binance) as a reference. The depth of liquidity in the price interval is visualized by the intensity of the shaded area; darker price intervals hold more liquidity than lighter price intervals at the same time. Liquidity is represented relatively at a point in time. Observe that the Binance price exits the price range for which trading is possible on Uniswap V3. } \label{fig:priceandliquidity}\vspace{-10pt}
\end{figure*}

To better understand the failure of Uniswap V3 to react to UST and USDT price fluctuations, as well as sudden price drops, we visualize the liquidity distribution over time in the four Uniswap V3 pools (cf. Figure~\ref{fig:priceandliquidity}). We notice that, as expected, the vast majority of the liquidity is distributed around the equal price of the pool's assets in our four pools. In the two liquidity pools that hold UST as one of their assets (cf. Figures~\ref{fig:priceandliquidityUSDC-UST-0.0001} and~\ref{fig:priceandliquidityUST-USDT-0.0005}), the relative liquidity distribution changes only very slightly over time. On 12 May, when the price of UST had not been near its peg for a couple of days, the pool's liquidity was still concentrated around the peg. In fact, by the end of 12 May, more than 99\% of liquidity was still on the price interval [0.99, 1.01] in both of the UST liquidity pools shown. Thus, it appears that liquidity providers are not adjusting their positions quickly enough, which explains why the pool's price did not track the primary market's price. 

In Figures~\ref{fig:priceandliquidityUSDC-USDT-0.0001} and~\ref{fig:priceandliquidityUSDC-USDT-0.0005}, we visualize the liquidity distribution in the two USDT-USDC Uniswap V3 pools and again notice that there are little to no changes in the liquidity distribution despite the price drop of USDT. Liquidity was not concentrated around the price but around USDT's intended peg. Thus, when the price moves outside the small interval around the equal price of the two stablecoins, there is little liquidity to support trading -- illustrating why the frequent sharp price drops in the Uniswap V3 pools were possible.  

\begin{figure*}[t]
\centering
  
    \begin{subfigure}[b]{0.48\linewidth}
          \includegraphics[scale =1,right]{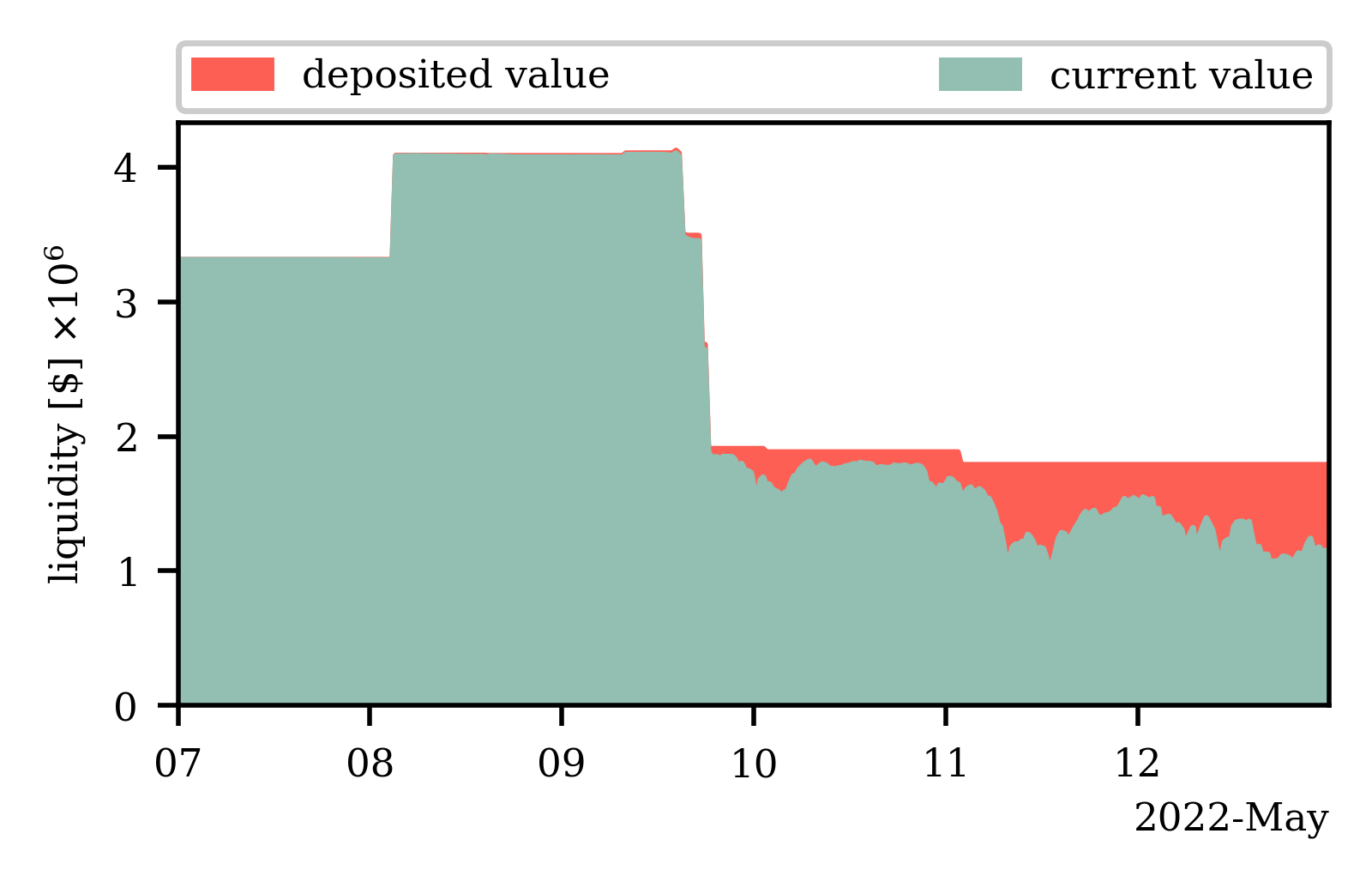}\vspace{-13pt}
        \caption{UST-USDC (f =0.01\%)} \label{fig:liquidityUSDC-UST-0.0001}
    \end{subfigure}%
    \hfill
    \begin{subfigure}[b]{0.48\linewidth}
        \includegraphics[scale =1,right]{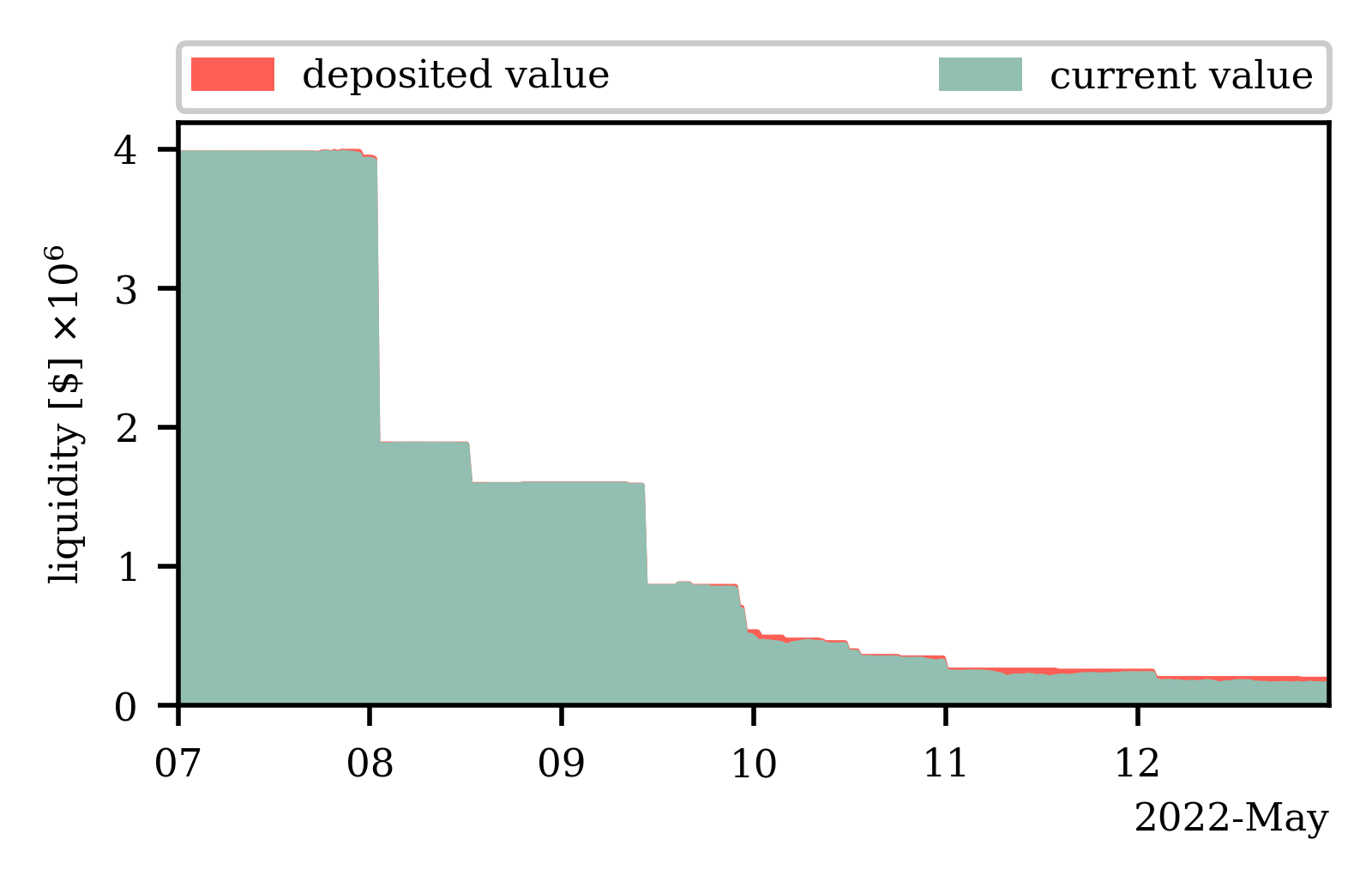}\vspace{-13pt}
        \caption{UST-USDC (f =0.05\%)} \label{fig:liquidityUSDC-UST-0.0005}
    \end{subfigure}\vspace{-3pt}
    
    \begin{subfigure}[b]{0.48\linewidth}
        \includegraphics[scale =1,right]{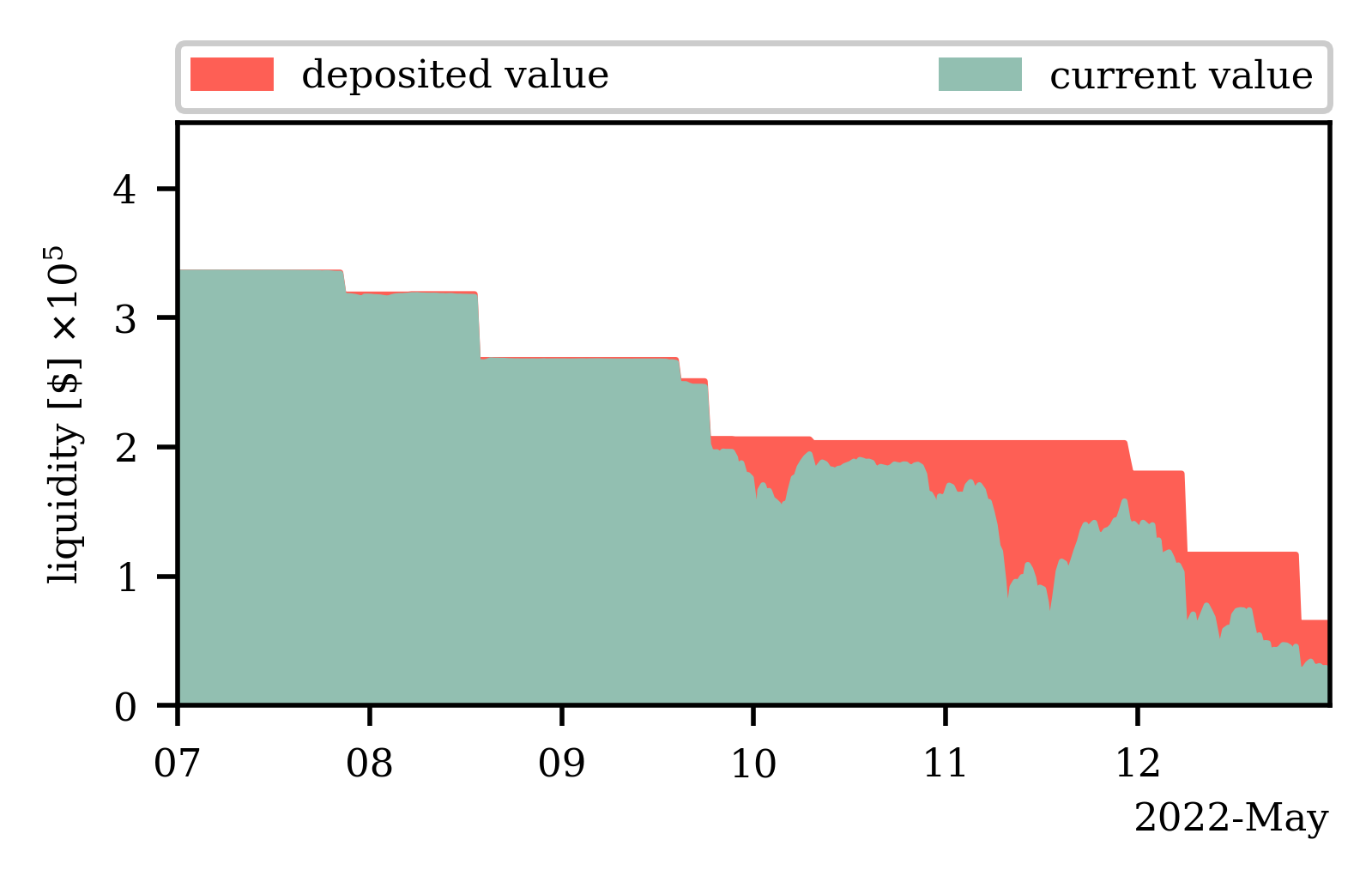}\vspace{-13pt}
        \caption{UST-USDT (f =0.05\%)} \label{fig:liquidityUST-USDT-0.0005}
    \end{subfigure}
    \hfill
    \begin{subfigure}[b]{0.48\linewidth}
        \includegraphics[scale =1,right]{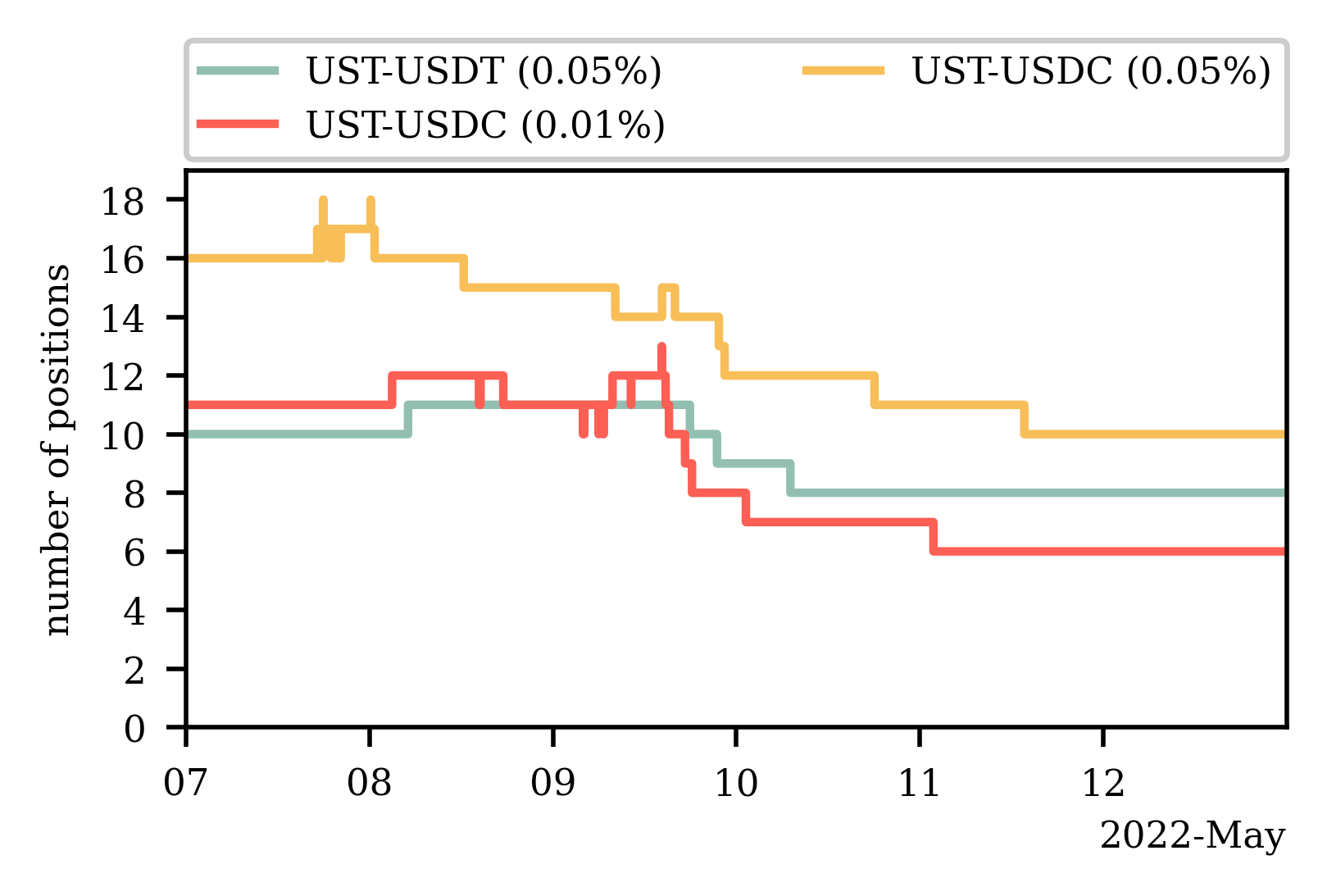}\vspace{-13pt}
        \caption{number of unique active liquidity positions} \label{fig:positions}
    \end{subfigure}
    \vspace{-8pt}
\caption{We visualize the liquidity levels over time in three Uniswap V3 pools: UST-USDC (f =0.01\%) (cf. Figure~\ref{fig:liquidityUSDC-UST-0.0001}), UST-USDC (f =0.05\%) (cf. Figure~\ref{fig:liquidityUSDC-UST-0.0005}) and UST-USDT (f =0.05\%) (cf. Figure~\ref{fig:liquidityUST-USDT-0.0005}). We plot both the original and current US\$ value of the liquidity that remains in the pool. Note that a pool's liquidity does not match the total value locked in the pool, as a consequence of the collected fees not acting as liquidity. In Figure~\ref{fig:positions} we further plot the number of active liquidity positions in the three pools over time. Notice that the number of unique liquidity positions is small and does not even half during UST price decline.} \label{fig:liquidityUST}\vspace{-10pt}
\end{figure*}

While the previous analysis outlined that the liquidity distribution across the price interval only saw very few adjustments in the face of dramatic price drops, even after a few days, we analyze whether and how quickly liquidity was removed from the Uniswap V3 stablecoin pools that held UST as one of their assets in Figure~\ref{fig:liquidityUST}. In Figures~\ref{fig:liquidityUSDC-UST-0.0001},~\ref{fig:liquidityUSDC-UST-0.0005} and~\ref{fig:liquidityUST-USDT-0.0005} we visualize the liquidity levels in three Uniswap V3 pools: UST-USDC (f =0.01\%), UST-USDC (f =0.05\%) and UST-USDT (f =0.05\%). Note that we plot both the value of the liquidity when it was deposited (in green) and the current value of the liquidity (in red) in US\$ for each timestamp. The current value of the liquidity accounts for UST's price drop. In all three pools, we observe that only a couple of significant liquidity providers remove their position from the liquidity pools before UST plummets below 0.95 cents on 9 May (cf. Figure~\ref{fig:priceacc}). This is especially apparent in the UST-USDC (f =0.05\%) pool, which we show in Figure~\ref{fig:liquidityUSDC-UST-0.0005}. Three large positions with an average value of 1M US\$, accounting for three-fourths of the pool's liquidity, are removed from the pool by midday on 9 May. From that point on, we see the difference between the value deposited and current begin to divide significantly, i.e., the liquidity providers in the pool start to experience significant losses. We further note that the liquidity providers were no longer earning fees, as the pools had stopped trading (cf. Figure~\ref{fig:priceacc}). Observe that even by the end of 12 May, by that time UST had not been trading near its peg for several days, more than half of the pool's liquidity positions were still active (cf. Figure~\ref{fig:positions}). We conclude that the small number of liquidity providers in the three pools did not adapt to the changes in the price of the pool's assets quickly enough to ensure that trading could continue or even to remove their assets from the pool in time to avoid significant financial losses. 

\section{Discussion}
Uniswap V3 does not appear equipped to handle volatile cryptocurrency markets. Its Achilles heel partially lies in its reliance on liquidity providers to react to market changes. While market making has long been handled by a few professional market makers in traditional finance, we observed that Uniswap V3 liquidity providers as a whole currently lack the sophistication and agility required for complex market making. The liquidity providers' apparent inexperience calls into question the suitability of the concentrated liquidity CPMM for DeFi. Not only did we observe that Uniswap V3 could not track the price of the primary market, but also saw that both its predecessor and Curve were fit for the task. While Uniswap V2 does not offer the same capital efficiency that Uniswap V3 for stablecoin pairs that continuously trade at an equal price, Curve's price curve can also reach high levels of capital efficiency around the equal price. 

Providing liquidity bears the risk of seeing the value of the liquidity position decrease compared to the initial assets deposited. This phenomenon is known as \emph{impermanent loss} and is particularly pronounced for high price volatility~\cite{Heimbach2022risks}. Given the speed of such an abrupt price drop and the general risk-off market sentiment that prevails in such a crisis, liquidity providers may be reluctant to open new liquidity positions.

Furthermore, note that pairs deemed to have a stable price, namely stablecoin pairs, generally have all liquidity in pools with a small percentage fee. Thus, once the price of a supposedly stable cryptocurrency pair becomes volatile, there are too few incentives for liquidity providers to create new liquidity positions. As a pool's fee cannot be adjusted, the meager fees liquidity providers expect to receive in the pools do not compensate them for the risk they face given the elevated volatility. Therefore, before supplying liquidity, they might choose to create a new liquidity pool. In fact, new pools that hold UST with high fees were launched after the price drop, such as the UST-USDC ($f=1\%$) pool. However, the pool was only created on 12 May, days after UST had lost its peg and never acquired the same levels of liquidity as the existing pools with lower fees. Thus, liquidity was further diluted across several pools, which reduces capital efficiency. 

Given this unfavorable risk-reward ratio for liquidity providers to deposit liquidity around the new price during times of market distress, DEXes could compensate liquidity providers for taking on this risk. In fact, liquidity providing on Uniswap V3 has been shown to exhibit similarities with financial derivatives ~\cite{Heimbach2022risks}. Thus, just as option prices rise with volatility~\cite{BlackScholes}, protocols could adopt a fee structure that would increase with volatility. Such behavior also occurs in traditional financial markets. During times of market upheaval the bid-ask spread increases as market makers, who generate their profit from this spread, seek compensation for the risks arising due to the rapid price changes~\cite{2022bidask}.

Finally, we also want to emphasize that the relative importance we should place on capital efficiency, in comparison to price accuracy, is unclear. The security implications of price inaccuracies and low levels of liquidity outside a small price interval around the equal price could allow for an inexpensive attack geometric mean time-weighted average price (TWAP) on Uniswap V3~\cite{Bentley2021Manipulating}. During such an attack, the attacker profits from manipulating Uniswap V3's spot price for the duration of a block. Not only do we already observe repeated price drops in the USDT-USDC (f = 0.05\%) pool, but we also found that in the UST pool, very little liquidity was outside the price interval [0.99,1.01] around the equal price at the end of 12 May. To give an example, on May 12 a single transaction exchanged 999'242 USDT for 995'221 USDC in the USDT-USDC (f = 0.01\%) pool. Thus, the trade executed at the price of 0.9959 (USDC/USDT). In the process, the trade moved the pool's price from  0.9993 to 0.4999. At the same time, the price on Binance was 0.9966, and we see that the trade almost executed at market price but created a huge price inaccuracy as it entered regions on the price interval with little liquidity. Thus, it was relatively inexpensive for an attacker to move the price.


\section{Conclusion}

Our empirical analysis finds that the concentrated liquidity CPMM pioneered by Uniswap V3 is currently not ready to handle unexpected price drops. Not only did we observe significant price inaccuracies and financial loss on the liquidity provider side, but we also see the potential for TWAP attacks on Uniswap V3 oracles. For the success of Uniswap V3, it is thus imperative that liquidity providers will become more sophisticated and agile.

\bibliographystyle{ACM-Reference-Format}
\bibliography{references}



\end{document}

%% file: TikzFiles/virtual.tikz
\definecolor{color1}{HTML}{92BFB1}
\definecolor{color2}{HTML}{FE5F55}
\definecolor{color3}{HTML}{f8be57}
\definecolor{color4}{HTML}{83B692}  

  
  
  
  
  

\begin{tikzpicture}[scale=1.5]
  \draw[-stealth, thick] (0, 0) -- (3.5, 0) node[midway, below,sloped,inner sep = 7 pt] {\small $X$ reserves};
  \draw[-stealth, thick] (0, 0) -- (0, 3.5) node[midway, above,sloped,inner sep = 7 pt] {\small $Y$ reserves};
  \draw[scale=1, line width=0.4mm, domain=0.57142:3.5, smooth, variable=\x, color2] plot ({\x}, {2/\x});
  \draw[scale=1, line width=0.4mm,color3] (0.25030348986165374,3.49)--(0.2523239188078164,3.48)--(0.25433802157220003,3.47)--(0.2563457906287525,3.46)--(0.2583472184394786,3.45)--(0.26,3.4417192735328204)--(0.260365687259121,3.4399999999999995)--(0.2624903704984537,3.43)--(0.26460836996759696,3.42)--(0.2667196771541789,3.41)--(0.2688242835313669,3.4)--(0.27,3.394401948630896)--(0.2709822345331502,3.39)--(0.2732096836626371,3.38)--(0.27543009133102253,3.37)--(0.2776434479652634,3.36)--(0.2798497439749554,3.3499999999999996)--(0.28,3.3493183901711987)--(0.2821763224739789,3.34)--(0.28450484722089564,3.33)--(0.2868259729335761,3.3200000000000003)--(0.28913968890245295,3.31)--(0.29,3.3062756732568794)--(0.29153183951832295,3.3)--(0.2939673122932644,3.29)--(0.2963950406094409,3.2800000000000002)--(0.2988150125677564,3.2699999999999996)--(0.3,3.265093905763969)--(0.3012969107051935,3.2600000000000002)--(0.3038379947707446,3.25)--(0.30637099179915317,3.24)--(0.3088958886312062,3.2300000000000004)--(0.30999999999999994,3.2256193045326933)--(0.31148949782620705,3.22)--(0.31413470576749847,3.21)--(0.3167714871344671,3.2)--(0.31939982743699796,3.19)--(0.32,3.187713693383634)--(0.32212501239593583,3.18)--(0.324872722147162,3.17)--(0.32761166895244664,3.16)--(0.33,3.151254885276365)--(0.3303589299588687,3.15)--(0.33321647860218023,3.1400000000000006)--(0.336064948473079,3.13)--(0.3389043222505104,3.12)--(0.34,3.1161348088627268)--(0.3418179041094653,3.1100000000000003)--(0.34477469396755905,3.1)--(0.3477220766656087,3.09)--(0.35000000000000003,3.082251058540428)--(0.35069049871170926,3.08)--(0.3537542985823906,3.0700000000000003)--(0.35680838655114655,3.06)--(0.35985274220612673,3.0500000000000003)--(0.36,3.049515929175602)--(0.3630156103672371,3.04)--(0.3661750051763887,3.0300000000000002)--(0.3693243677271707,3.02)--(0.37,3.01785219047826)--(0.37256902185812246,3.0100000000000002)--(0.37583224156655626,3.0)--(0.3790851354556876,2.99)--(0.38,2.9871837890976467)--(0.38242355881910073,2.98)--(0.3857890485373461,2.9700000000000006)--(0.3891439252480349,2.96)--(0.38999999999999996,2.957445068393972)--(0.39258696754693057,2.95)--(0.39605310930186555,2.94)--(0.3995083571820315,2.93)--(0.4,2.928575761094741)--(0.4030658002394427,2.92)--(0.4066309221557821,2.91)--(0.41000000000000003,2.900521468171386)--(0.4101917070430878,2.9)--(0.4138654980163686,2.89)--(0.4175278828336513,2.88)--(0.42,2.8732344697587173)--(0.42122091152209107,2.87)--(0.42499047126136125,2.86)--(0.4287483642085091,2.8499999999999996)--(0.43,2.8466646465929433)--(0.4325806393804156,2.84)--(0.43644417702588056,2.83)--(0.44,2.8207697946820844)--(0.4403056565845534,2.82)--(0.44427349199672894,2.81)--(0.4482291932997911,2.8000000000000003)--(0.45000000000000007,2.7955159517171455)--(0.4522428473183072,2.79)--(0.4563012565646709,2.7800000000000002)--(0.46,2.770860359866922)--(0.4603581333260251,2.77)--(0.4645178396115245,2.7600000000000002)--(0.46866497426814024,2.7500000000000004)--(0.47000000000000003,2.7467766358233625)--(0.4728841934593305,2.74)--(0.47713099973009454,2.73)--(0.48,2.723229499836975)--(0.48140499856759583,2.72)--(0.48575002437180914,2.71)--(0.49,2.700189921618631)--(0.4900843991456369,2.7)--(0.4945261749542298,2.69)--(0.4989547987828313,2.68)--(0.5,2.6776374536790044)--(0.5034630800006102,2.67)--(0.5079868088538604,2.66)--(0.51,2.655542816829101)--(0.512563901741284,2.65)--(0.517181253114418,2.64)--(0.52,2.633883420838944)--(0.5218313669581265,2.63)--(0.5265408536286772,2.62)--(0.53,2.6126384777167706)--(0.5312677966779333,2.61)--(0.5360679303511701,2.5999999999999996)--(0.54,2.591788491260455)--(0.5408751346792358,2.59)--(0.5457644293322844,2.58)--(0.55,2.57131515552952)--(0.5506549747905447,2.57)--(0.555631949732252,2.56)--(0.56,2.551201262540192)--(0.560608586969781,2.5500000000000003)--(0.565671769633169,2.54)--(0.5700000000000001,2.5314306182073154)--(0.5707369421648542,2.5300000000000002)--(0.5758848706486883,2.52)--(0.58,2.511987965672389)--(0.581040735964111,2.5100000000000002)--(0.5862719613397017,2.5)--(0.59,2.492858915257225)--(0.5915204110528379,2.49)--(0.596833499451717,2.48)--(0.6,2.4740298803699936)--(0.602176178498258,2.47)--(0.6075697129958507,2.46)--(0.6100000000000001,2.4554880187664843)--(0.6130080378906764,2.45)--(0.6184806202005174,2.44)--(0.62,2.437221178635895)--(0.6240157963726632,2.43)--(0.6295660483651173,2.42)--(0.63,2.4192178490386667)--(0.635199086591566,2.41)--(0.64,2.401470508047546)--(0.6408410159770732,2.4)--(0.64655738361328,2.39)--(0.65,2.383967854265074)--(0.652299922452634,2.38)--(0.6580900208371929,2.37)--(0.66,2.366698246214175)--(0.663933666240357,2.36)--(0.6697962049536195,2.35)--(0.67,2.34965231587449)--(0.6757413156202425,2.34)--(0.68,2.332827857425858)--(0.6817032307302578,2.33)--(0.6877218386581957,2.32)--(0.69,2.316211125220131)--(0.6937867408283135,2.31)--(0.6998741156410174,2.3000000000000003)--(0.7,2.2997931887846317)--(0.7060419969034637,2.29)--(0.71,2.2835749757252124)--(0.712231361114939,2.2800000000000002)--(0.718467709623326,2.2699999999999996)--(0.72,2.267541557896264)--(0.7247608713967063,2.2600000000000002)--(0.73,2.2516895019478396)--(0.7310784070541325,2.25)--(0.7374592089143917,2.24)--(0.74,2.2360145170348193)--(0.7438809670046538,2.23)--(0.75,2.220505369817104)--(0.7503295513040851,2.22)--(0.7568504623497918,2.21)--(0.76,2.2051650573243045)--(0.7634033623954485,2.2)--(0.7699853564390337,2.1899999999999995)--(0.7700000000000001,2.1899777551190795)--(0.7766420164563121,2.18)--(0.78,2.174950355732821)--(0.7833279439185503,2.17)--(0.79,2.160065559510283)--(0.7900444989044461,2.16)--(0.7968333357558853,2.15)--(0.8,2.1453313066479525)--(0.8036536148092582,2.14)--(0.81,2.130731535308065)--(0.8105059640776089,2.13)--(0.8174229996202232,2.12)--(0.8200000000000001,2.1162721121759396)--(0.8243784818263935,2.11)--(0.8300000000000001,2.1019413380383636)--(0.8313672264595697,2.1)--(0.8384088777557821,2.09)--(0.8399999999999999,2.0877399223796167)--(0.8455000214019444,2.08)--(0.8499999999999999,2.07366347007003)--(0.8526254051583102,2.07)--(0.8597882863506895,2.0600000000000005)--(0.86,2.059704518092684)--(0.8670152273149273,2.05)--(0.87,2.045868925739346)--(0.8742772037923705,2.04)--(0.8799999999999999,2.0321439482923993)--(0.881574988980235,2.0300000000000002)--(0.8889207418101237,2.02)--(0.8899999999999999,2.0185309265312847)--(0.8963190343161173,2.0100000000000002)--(0.9,2.005028853218379)--(0.903753690322149,2.0)--(0.91,1.9916292936762146)--(0.9112253476381005,1.99)--(0.9187471268754633,1.98)--(0.92,1.9783345770455398)--(0.9263180844794104,1.9699999999999998)--(0.9300000000000002,1.9651420528416081)--(0.9339264127893948,1.96)--(0.9400000000000001,1.952044891512787)--(0.9415726396930166,1.95)--(0.9492642099139285,1.94)--(0.9500000000000001,1.9390437575685509)--(0.9570084491824842,1.93)--(0.96,1.9261395602931335)--(0.9647908181803897,1.92)--(0.97,1.9133243981262522)--(0.9726117567855262,1.9100000000000001)--(0.98,1.900596382802675)--(0.980471673838654,1.9000000000000001)--(0.9883851357608826,1.8900000000000001)--(0.9899999999999999,1.8879598754912157)--(0.9963414849486223,1.88)--(1.0,1.8754085812257097)--(1.0043369075839914,1.87)--(1.01,1.8629390359309292)--(1.0123717491059194,1.8600000000000003)--(1.02,1.8505496295287556)--(1.0204463314293934,1.85)--(1.0285726466420066,1.84)--(1.03,1.838244328138547)--(1.0367421142675952,1.83)--(1.04,1.8260176856937975)--(1.0449513402194244,1.82)--(1.05,1.8138665617277598)--(1.0532006032609487,1.81)--(1.06,1.8017895801902966)--(1.061490164862914,1.8)--(1.0698216371519005,1.7899999999999998)--(1.07,1.789786115883435)--(1.0782046736255029,1.7800000000000002)--(1.08,1.777859840874893)--(1.0866279968550043,1.77)--(1.09,1.766003740786569)--(1.095091841922758,1.7599999999999998)--(1.1,1.7542166374930461)--(1.103596431816571,1.75)--(1.11,1.7424973969309892)--(1.1121419786124265,1.74)--(1.12,1.730844927440014)--(1.1207286846082736,1.73)--(1.1293611877124186,1.72)--(1.1300000000000001,1.7192607380886331)--(1.1380397614136226,1.71)--(1.1400000000000001,1.707744011179906)--(1.146759492367531,1.6999999999999997)--(1.1500000000000001,1.6962908809213082)--(1.1555205767833874,1.69)--(1.16,1.6849004111784744)--(1.1643232051282757,1.68)--(1.17,1.6735717007707571)--(1.1731675630564156,1.67)--(1.18,1.6623038822092924)--(1.1820538323038439,1.6600000000000001)--(1.19,1.651096120488248)--(1.1909821915505243,1.6500000000000001)--(1.1999531108828019,1.6400000000000001)--(1.2,1.6399478062198471)--(1.208972225336951,1.6300000000000001)--(1.21,1.6288618575723817)--(1.2180335220071032,1.62)--(1.22,1.6178334998385027)--(1.2271371919833405,1.61)--(1.23,1.6068620162257274)--(1.2362834264499685,1.6)--(1.2399999999999998,1.595946716860782)--(1.245472417420822,1.59)--(1.25,1.5850869378400119)--(1.2547043584533801,1.58)--(1.26,1.5742820403144497)--(1.263979445343327,1.57)--(1.27,1.5635314096074415)--(1.2732978768011514,1.56)--(1.28,1.5528344543628059)--(1.2826598551123307,1.55)--(1.29,1.542190605721608)--(1.2920655867826167,1.54)--(1.3,1.5315993165256963)--(1.301515283169889,1.53)--(1.31,1.5210600605462234)--(1.3110091611040204,1.52)--(1.32,1.510572331735447)--(1.3205474434961613,1.51)--(1.33,1.5001356435001574)--(1.330130359938826,1.5)--(1.3397593923879236,1.49)--(1.34,1.4897506089631403)--(1.349433942269899,1.48)--(1.3500000000000003,1.47941609392748)--(1.359153591999011,1.4700000000000002)--(1.36,1.469131082780413)--(1.3689186123667714,1.46)--(1.37,1.4588951555451275)--(1.378729283325554,1.4499999999999997)--(1.3800000000000001,1.4487079080387528)--(1.3885858945765575,1.44)--(1.3900000000000001,1.4385689512348885)--(1.3984887461569258,1.43)--(1.4000000000000001,1.4284779106369763)--(1.408438149027312,1.42)--(1.41,1.4184344256611594)--(1.4184344256611594,1.41)--(1.42,1.408438149027312)--(1.4284779106369763,1.4000000000000001)--(1.43,1.3984887461569258)--(1.4385689512348885,1.3900000000000001)--(1.44,1.3885858945765575)--(1.4487079080387528,1.3800000000000001)--(1.4499999999999997,1.378729283325554)--(1.4588951555451275,1.37)--(1.46,1.3689186123667714)--(1.469131082780413,1.36)--(1.4700000000000002,1.359153591999011)--(1.47941609392748,1.3500000000000003)--(1.48,1.349433942269899)--(1.4897506089631403,1.34)--(1.49,1.3397593923879236)--(1.5,1.330130359938826)--(1.5001356435001574,1.33)--(1.51,1.3205474434961613)--(1.510572331735447,1.32)--(1.52,1.3110091611040204)--(1.5210600605462234,1.31)--(1.53,1.301515283169889)--(1.5315993165256963,1.3)--(1.54,1.2920655867826167)--(1.542190605721608,1.29)--(1.55,1.2826598551123307)--(1.5528344543628059,1.28)--(1.56,1.2732978768011514)--(1.5635314096074415,1.27)--(1.57,1.263979445343327)--(1.5742820403144497,1.26)--(1.58,1.2547043584533801)--(1.5850869378400119,1.25)--(1.59,1.245472417420822)--(1.595946716860782,1.2399999999999998)--(1.6,1.2362834264499685)--(1.6068620162257274,1.23)--(1.61,1.2271371919833405)--(1.6178334998385027,1.22)--(1.62,1.2180335220071032)--(1.6288618575723817,1.21)--(1.6300000000000001,1.208972225336951)--(1.6399478062198471,1.2)--(1.6400000000000001,1.1999531108828019)--(1.6500000000000001,1.1909821915505243)--(1.651096120488248,1.19)--(1.6600000000000001,1.1820538323038439)--(1.6623038822092924,1.18)--(1.67,1.1731675630564156)--(1.6735717007707571,1.17)--(1.68,1.1643232051282757)--(1.6849004111784744,1.16)--(1.69,1.1555205767833874)--(1.6962908809213082,1.1500000000000001)--(1.6999999999999997,1.146759492367531)--(1.707744011179906,1.1400000000000001)--(1.71,1.1380397614136226)--(1.7192607380886331,1.1300000000000001)--(1.72,1.1293611877124186)--(1.73,1.1207286846082736)--(1.730844927440014,1.12)--(1.74,1.1121419786124265)--(1.7424973969309892,1.11)--(1.75,1.103596431816571)--(1.7542166374930461,1.1)--(1.7599999999999998,1.095091841922758)--(1.766003740786569,1.09)--(1.77,1.0866279968550043)--(1.777859840874893,1.08)--(1.7800000000000002,1.0782046736255029)--(1.789786115883435,1.07)--(1.7899999999999998,1.0698216371519005)--(1.8,1.061490164862914)--(1.8017895801902966,1.06)--(1.81,1.0532006032609487)--(1.8138665617277598,1.05)--(1.82,1.0449513402194244)--(1.8260176856937975,1.04)--(1.83,1.0367421142675952)--(1.838244328138547,1.03)--(1.84,1.0285726466420066)--(1.85,1.0204463314293934)--(1.8505496295287556,1.02)--(1.8600000000000003,1.0123717491059194)--(1.8629390359309292,1.01)--(1.87,1.0043369075839914)--(1.8754085812257097,1.0)--(1.88,0.9963414849486223)--(1.8879598754912157,0.9899999999999999)--(1.8900000000000001,0.9883851357608826)--(1.9000000000000001,0.980471673838654)--(1.900596382802675,0.98)--(1.9100000000000001,0.9726117567855262)--(1.9133243981262522,0.97)--(1.92,0.9647908181803897)--(1.9261395602931335,0.96)--(1.93,0.9570084491824842)--(1.9390437575685509,0.9500000000000001)--(1.94,0.9492642099139285)--(1.95,0.9415726396930166)--(1.952044891512787,0.9400000000000001)--(1.96,0.9339264127893948)--(1.9651420528416081,0.9300000000000002)--(1.9699999999999998,0.9263180844794104)--(1.9783345770455398,0.92)--(1.98,0.9187471268754633)--(1.99,0.9112253476381005)--(1.9916292936762146,0.91)--(2.0,0.903753690322149)--(2.005028853218379,0.9)--(2.0100000000000002,0.8963190343161173)--(2.0185309265312847,0.8899999999999999)--(2.02,0.8889207418101237)--(2.0300000000000002,0.881574988980235)--(2.0321439482923993,0.8799999999999999)--(2.04,0.8742772037923705)--(2.045868925739346,0.87)--(2.05,0.8670152273149273)--(2.059704518092684,0.86)--(2.0600000000000005,0.8597882863506895)--(2.07,0.8526254051583102)--(2.07366347007003,0.8499999999999999)--(2.08,0.8455000214019444)--(2.0877399223796167,0.8399999999999999)--(2.09,0.8384088777557821)--(2.1,0.8313672264595697)--(2.1019413380383636,0.8300000000000001)--(2.11,0.8243784818263935)--(2.1162721121759396,0.8200000000000001)--(2.12,0.8174229996202232)--(2.13,0.8105059640776089)--(2.130731535308065,0.81)--(2.14,0.8036536148092582)--(2.1453313066479525,0.8)--(2.15,0.7968333357558853)--(2.16,0.7900444989044461)--(2.160065559510283,0.79)--(2.17,0.7833279439185503)--(2.174950355732821,0.78)--(2.18,0.7766420164563121)--(2.1899777551190795,0.7700000000000001)--(2.1899999999999995,0.7699853564390337)--(2.2,0.7634033623954485)--(2.2051650573243045,0.76)--(2.21,0.7568504623497918)--(2.22,0.7503295513040851)--(2.220505369817104,0.75)--(2.23,0.7438809670046538)--(2.2360145170348193,0.74)--(2.24,0.7374592089143917)--(2.25,0.7310784070541325)--(2.2516895019478396,0.73)--(2.2600000000000002,0.7247608713967063)--(2.267541557896264,0.72)--(2.2699999999999996,0.718467709623326)--(2.2800000000000002,0.712231361114939)--(2.2835749757252124,0.71)--(2.29,0.7060419969034637)--(2.2997931887846317,0.7)--(2.3000000000000003,0.6998741156410174)--(2.31,0.6937867408283135)--(2.316211125220131,0.69)--(2.32,0.6877218386581957)--(2.33,0.6817032307302578)--(2.332827857425858,0.68)--(2.34,0.6757413156202425)--(2.34965231587449,0.67)--(2.35,0.6697962049536195)--(2.36,0.663933666240357)--(2.366698246214175,0.66)--(2.37,0.6580900208371929)--(2.38,0.652299922452634)--(2.383967854265074,0.65)--(2.39,0.64655738361328)--(2.4,0.6408410159770732)--(2.401470508047546,0.64)--(2.41,0.635199086591566)--(2.4192178490386667,0.63)--(2.42,0.6295660483651173)--(2.43,0.6240157963726632)--(2.437221178635895,0.62)--(2.44,0.6184806202005174)--(2.45,0.6130080378906764)--(2.4554880187664843,0.6100000000000001)--(2.46,0.6075697129958507)--(2.47,0.602176178498258)--(2.4740298803699936,0.6)--(2.48,0.596833499451717)--(2.49,0.5915204110528379)--(2.492858915257225,0.59)--(2.5,0.5862719613397017)--(2.5100000000000002,0.581040735964111)--(2.511987965672389,0.58)--(2.52,0.5758848706486883)--(2.5300000000000002,0.5707369421648542)--(2.5314306182073154,0.5700000000000001)--(2.54,0.565671769633169)--(2.5500000000000003,0.560608586969781)--(2.551201262540192,0.56)--(2.56,0.555631949732252)--(2.57,0.5506549747905447)--(2.57131515552952,0.55)--(2.58,0.5457644293322844)--(2.59,0.5408751346792358)--(2.591788491260455,0.54)--(2.5999999999999996,0.5360679303511701)--(2.61,0.5312677966779333)--(2.6126384777167706,0.53)--(2.62,0.5265408536286772)--(2.63,0.5218313669581265)--(2.633883420838944,0.52)--(2.64,0.517181253114418)--(2.65,0.512563901741284)--(2.655542816829101,0.51)--(2.66,0.5079868088538604)--(2.67,0.5034630800006102)--(2.6776374536790044,0.5)--(2.68,0.4989547987828313)--(2.69,0.4945261749542298)--(2.7,0.4900843991456369)--(2.700189921618631,0.49)--(2.71,0.48575002437180914)--(2.72,0.48140499856759583)--(2.723229499836975,0.48)--(2.73,0.47713099973009454)--(2.74,0.4728841934593305)--(2.7467766358233625,0.47000000000000003)--(2.7500000000000004,0.46866497426814024)--(2.7600000000000002,0.4645178396115245)--(2.77,0.4603581333260251)--(2.770860359866922,0.46)--(2.7800000000000002,0.4563012565646709)--(2.79,0.4522428473183072)--(2.7955159517171455,0.45000000000000007)--(2.8000000000000003,0.4482291932997911)--(2.81,0.44427349199672894)--(2.82,0.4403056565845534)--(2.8207697946820844,0.44)--(2.83,0.43644417702588056)--(2.84,0.4325806393804156)--(2.8466646465929433,0.43)--(2.8499999999999996,0.4287483642085091)--(2.86,0.42499047126136125)--(2.87,0.42122091152209107)--(2.8732344697587173,0.42)--(2.88,0.4175278828336513)--(2.89,0.4138654980163686)--(2.9,0.4101917070430878)--(2.900521468171386,0.41000000000000003)--(2.91,0.4066309221557821)--(2.92,0.4030658002394427)--(2.928575761094741,0.4)--(2.93,0.3995083571820315)--(2.94,0.39605310930186555)--(2.95,0.39258696754693057)--(2.957445068393972,0.38999999999999996)--(2.96,0.3891439252480349)--(2.9700000000000006,0.3857890485373461)--(2.98,0.38242355881910073)--(2.9871837890976467,0.38)--(2.99,0.3790851354556876)--(3.0,0.37583224156655626)--(3.0100000000000002,0.37256902185812246)--(3.01785219047826,0.37)--(3.02,0.3693243677271707)--(3.0300000000000002,0.3661750051763887)--(3.04,0.3630156103672371)--(3.049515929175602,0.36)--(3.0500000000000003,0.35985274220612673)--(3.06,0.35680838655114655)--(3.0700000000000003,0.3537542985823906)--(3.08,0.35069049871170926)--(3.082251058540428,0.35000000000000003)--(3.09,0.3477220766656087)--(3.1,0.34477469396755905)--(3.1100000000000003,0.3418179041094653)--(3.1161348088627268,0.34)--(3.12,0.3389043222505104)--(3.13,0.336064948473079)--(3.1400000000000006,0.33321647860218023)--(3.15,0.3303589299588687)--(3.151254885276365,0.33)--(3.16,0.32761166895244664)--(3.17,0.324872722147162)--(3.18,0.32212501239593583)--(3.187713693383634,0.32)--(3.19,0.31939982743699796)--(3.2,0.3167714871344671)--(3.21,0.31413470576749847)--(3.22,0.31148949782620705)--(3.2256193045326933,0.30999999999999994)--(3.2300000000000004,0.3088958886312062)--(3.24,0.30637099179915317)--(3.25,0.3038379947707446)--(3.2600000000000002,0.3012969107051935)--(3.265093905763969,0.3)--(3.2699999999999996,0.2988150125677564)--(3.2800000000000002,0.2963950406094409)--(3.29,0.2939673122932644)--(3.3,0.29153183951832295)--(3.3062756732568794,0.29)--(3.31,0.28913968890245295)--(3.3200000000000003,0.2868259729335761)--(3.33,0.28450484722089564)--(3.34,0.2821763224739789)--(3.3493183901711987,0.28)--(3.3499999999999996,0.2798497439749554)--(3.36,0.2776434479652634)--(3.37,0.27543009133102253)--(3.38,0.2732096836626371)--(3.39,0.2709822345331502)--(3.394401948630896,0.27)--(3.4,0.2688242835313669)--(3.41,0.2667196771541789)--(3.42,0.26460836996759696)--(3.43,0.2624903704984537)--(3.4399999999999995,0.260365687259121)--(3.4417192735328204,0.26)--(3.45,0.2583472184394786)--(3.46,0.2563457906287525)--(3.47,0.25433802157220003)--(3.48,0.2523239188078164)--(3.49,0.25030348986165374);
  

   \node at (2.857,0.7)[circle,fill,inner sep=1.3pt](A) {} ;
   \node at (2.857,1.41421)[](A1) {} ;
   
  \node at (2.857,0.7)[anchor = south east](F) {\small $l$} ;

   \node at (.7,2.857)[circle,fill,inner sep=1.3pt](B) {} ;
   \node at (1.41421,2.857)[](B1) {} ;
  
   \draw [stealth-stealth, line width=0.2mm] (.7,2.957) -- (1.41421,2.957) node[midway,above] {\small$x_{\text{real}}$}  ;
   \node at (0.7,2.857)[anchor = north west](E) {\small $u$} ;
  \node at (1.41421,1.41421)[circle,fill,inner sep=1.3pt](C) {} ;

  \node at (1.41421,1.41421)[anchor = north east](G) {\small $m$} ;
  \draw [dotted] (A) -- (A1.center);
  \draw [dotted] (A1.center) -- (C);
  \draw [dotted] (B) -- (B1.center);
  \draw [dotted] (B1.center) -- (C);

  \path[line width=0.4mm,color2]    (2.06,3.3) edge	node {}	(2.27,3.3);
\path[ line width=0.4mm,color3]     (2.06,3.1) edge	node {}	(2.27,3.1);
    \node[anchor=west](v6) at  (2.34,3.3) {\small CPMM};
  \node[anchor=west](v7) at  (2.34,3.1) {\small Curve CFMM};
  \draw[line width=0.4mm,rounded corners,color = gray,opacity=0.5] (2, 2.97) rectangle (3.52, 3.42) {};
  \draw [stealth-stealth, line width=0.2mm] (2.957,0.7) -- (2.957,1.41421) node[midway,right] {\small$y_{\text{real}}$}  ;
  
  
  \draw [stealth-stealth, line width=0.2mm] (0,1.4) -- (1.37,1.41421) node[midway,above] {\small$x$}  ;
  
  \draw [stealth-stealth, line width=0.2mm] (1.4,0) -- (1.41421,1.37) node[midway,right] {\small$y$}  ;
  
\end{tikzpicture}

%% file: TikzFiles/liquidityV2.tikz
\definecolor{color1}{HTML}{F6AE2D}
\definecolor{color2}{HTML}{FE5F55}
\definecolor{color3}{HTML}{7180AC}
\definecolor{color4}{HTML}{83B692}  
\begin{tikzpicture}[scale=0.17]
\filldraw[fill=color2!20!white, draw=color2,line width=0.4mm] (0,0) rectangle (14,4);
  \draw[-stealth, thick] (0, 0) -- (15, 0) node[midway, below,sloped,inner sep = 7pt] {\small price};
  \draw[-stealth, thick] (0, 0) -- (0, 8) node[midway, above,sloped,label distance =1.2cm] {\small liquidity};
	\node at (0,-0.99) {\small $0$};
     \node at (14,-0.99) {\small $\infty$};
\end{tikzpicture}

%% file: TikzFiles/liquidityV3.tikz
\definecolor{color1}{HTML}{F6AE2D}
\definecolor{color2}{HTML}{FE5F55}
\definecolor{color3}{HTML}{7180AC}
\definecolor{color4}{HTML}{83B692}  

\begin{tikzpicture}[scale = 0.17]
\filldraw[fill=color2!20!white, draw=color2,  line width=0.4mm] (4,0) rectangle (12,7);
  \draw[-stealth, thick] (0, 0) -- (16, 0) node[midway, below,sloped,inner sep = 7pt] {\small price};
  \draw[-stealth, thick] (0, 0) -- (0, 8) node[midway, above,sloped,label distance =1.2cm] {\small liquidity};
	\node at (4,-0.99) {\small $P_l$};
     \node at (12,-0.99) {\small $P_u$};
\end{tikzpicture}

%% file: TikzFiles/liquiditys.tikz
\definecolor{color1}{HTML}{F6AE2D}
\definecolor{color2}{HTML}{FE5F55}
\definecolor{color3}{HTML}{7180AC}
\definecolor{color4}{HTML}{83B692}  
\begin{tikzpicture}[scale=0.17]
  \draw[-stealth, thick] (-3, 0) -- (-3, 11) node[midway, above,sloped,label distance =1.2cm] {\small liquidity};
	\node at (-2,-0.99) {\small $0$};
     \node at (20,-0.99) {\small $\infty$};

\filldraw[fill=color2!20!white, draw=color2, very thick] (8,0) rectangle (10,10);
\filldraw[fill=color2!20!white, draw=color2, very thick] (10,0) rectangle (12,9);
\filldraw[fill=color2!20!white, draw=color2, very thick] (-2,0) rectangle (0,0.5);
\filldraw[fill=color2!20!white, draw=color2, very thick] (0,0) rectangle (2,3);

\filldraw[fill=color2!20!white, draw=color2, very thick] (2,0) rectangle (4,4);
\filldraw[fill=color2!20!white, draw=color2, very thick] (6,0) rectangle (8,8);

\filldraw[fill=color2!20!white, draw=color2, very thick] (4,0) rectangle (6,7);

\filldraw[fill=color2!20!white, draw=color2, very thick] (16,0) rectangle (18,3);
\filldraw[fill=color2!20!white, draw=color2, very thick] (18,0) rectangle (20,1);

\filldraw[fill=color2!20!white, draw=color2, very thick] (14,0) rectangle (16,6);

\filldraw[fill=color2!20!white, draw=color2, very thick] (12,0) rectangle (14,7);
  \draw[-stealth, thick] (-3, 0) -- (21, 0) node[midway, below,sloped,inner sep = 7pt] {\small price};

\end{tikzpicture}

%% file: main.bbl

\begin{thebibliography}{26}


\ifx \showCODEN    \undefined \def \showCODEN     #1{\unskip}     \fi
\ifx \showDOI      \undefined \def \showDOI       #1{#1}\fi
\ifx \showISBNx    \undefined \def \showISBNx     #1{\unskip}     \fi
\ifx \showISBNxiii \undefined \def \showISBNxiii  #1{\unskip}     \fi
\ifx \showISSN     \undefined \def \showISSN      #1{\unskip}     \fi
\ifx \showLCCN     \undefined \def \showLCCN      #1{\unskip}     \fi
\ifx \shownote     \undefined \def \shownote      #1{#1}          \fi
\ifx \showarticletitle \undefined \def \showarticletitle #1{#1}   \fi
\ifx \showURL      \undefined \def \showURL       {\relax}        \fi
\providecommand\bibfield[2]{#2}
\providecommand\bibinfo[2]{#2}
\providecommand\natexlab[1]{#1}
\providecommand\showeprint[2][]{arXiv:#2}

\bibitem[202(2022a)]%
        {2021curve}
 \bibinfo{year}{2022}\natexlab{a}.
\newblock \bibinfo{title}{Curve}.
\newblock \bibinfo{howpublished}{\url{https://curve.fi/}}.
\newblock


\bibitem[202(2022b)]%
        {2022erigon}
 \bibinfo{year}{2022}\natexlab{b}.
\newblock \bibinfo{title}{Erigon}.
\newblock \bibinfo{howpublished}{\url{https://github.com/ledgerwatch/erigon}}.
\newblock


\bibitem[202(2022c)]%
        {2022bianancemarketdate}
 \bibinfo{year}{2022}\natexlab{c}.
\newblock \bibinfo{title}{Historical Market Data}.
\newblock
  \bibinfo{howpublished}{\url{https://www.binance.com/en/landing/data}}.
\newblock


\bibitem[202(2022d)]%
        {2022LUNA}
 \bibinfo{year}{2022}\natexlab{d}.
\newblock \bibinfo{title}{LUNA \& UST Trading Suspended}.
\newblock
  \bibinfo{howpublished}{\url{https://www.binance.com/en/support/announcement/f68451879a1841a6a0f44025735d9236}}.
\newblock


\bibitem[202(2022e)]%
        {2021sushiswap}
 \bibinfo{year}{2022}\natexlab{e}.
\newblock \bibinfo{title}{Sushiswap}.
\newblock \bibinfo{howpublished}{\url{https://sushi.com/}}.
\newblock


\bibitem[202(2022f)]%
        {2022terrausd}
 \bibinfo{year}{2022}\natexlab{f}.
\newblock \bibinfo{title}{Terrausd}.
\newblock
  \bibinfo{howpublished}{\url{https://coinmarketcap.com/de/currencies/terrausd/}}.
\newblock


\bibitem[202(2022g)]%
        {2022exchangeranking}
 \bibinfo{year}{2022}\natexlab{g}.
\newblock \bibinfo{title}{Top Cryptocurrency Spot Exchanges}.
\newblock
  \bibinfo{howpublished}{\url{https://coinmarketcap.com/rankings/exchanges/}}.
\newblock


\bibitem[202(2022h)]%
        {2022bidask}
 \bibinfo{year}{2022}\natexlab{h}.
\newblock \bibinfo{title}{Understanding bid ask spreads}.
\newblock
  \bibinfo{howpublished}{\url{https://www.rbcgam.com/en/ca/learn-plan/types-of-investments/understanding-bid-ask-spreads/detail}}.
\newblock


\bibitem[Adams et~al\mbox{.}(2020)]%
        {adams2020uniswap}
\bibfield{author}{\bibinfo{person}{Hayden Adams}, \bibinfo{person}{Noah
  Zinsmeister}, {and} \bibinfo{person}{Dan Robinson}.}
  \bibinfo{year}{2020}\natexlab{}.
\newblock \bibinfo{title}{Uniswap v2 Core}.
\newblock
\newblock


\bibitem[Adams et~al\mbox{.}(2021)]%
        {adams2021uniswap}
\bibfield{author}{\bibinfo{person}{Hayden Adams}, \bibinfo{person}{Noah
  Zinsmeister}, \bibinfo{person}{Moody Salem}, \bibinfo{person}{River Keefer},
  {and} \bibinfo{person}{Dan Robinson}.} \bibinfo{year}{2021}\natexlab{}.
\newblock \bibinfo{title}{Uniswap v3 core}.
\newblock
\newblock


\bibitem[Angeris and Chitra(2020)]%
        {angeris2020improved}
\bibfield{author}{\bibinfo{person}{Guillermo Angeris} {and}
  \bibinfo{person}{Tarun Chitra}.} \bibinfo{year}{2020}\natexlab{}.
\newblock \showarticletitle{Improved price oracles: Constant function market
  makers}. In \bibinfo{booktitle}{\emph{Proceedings of the 2nd ACM Conference
  on Advances in Financial Technologies}}.
\newblock


\bibitem[Angeris et~al\mbox{.}(2019)]%
        {angeris2019analysis}
\bibfield{author}{\bibinfo{person}{Guillermo Angeris},
  \bibinfo{person}{Hsien-Tang Kao}, \bibinfo{person}{Rei Chiang},
  \bibinfo{person}{Charlie Noyes}, {and} \bibinfo{person}{Tarun Chitra}.}
  \bibinfo{year}{2019}\natexlab{}.
\newblock \bibinfo{title}{An analysis of Uniswap markets}.
\newblock
\newblock


\bibitem[Bentley(2021)]%
        {Bentley2021Manipulating}
\bibfield{author}{\bibinfo{person}{Bentley}.} \bibinfo{year}{2021}\natexlab{}.
\newblock \bibinfo{title}{Manipulating Uniswap V3 TWAP Oracles}.
\newblock
\newblock


\bibitem[Berg et~al\mbox{.}(2022)]%
        {Berg2022empirical}
\bibfield{author}{\bibinfo{person}{Jan~Arvid Berg}, \bibinfo{person}{Robin
  Fritsch}, \bibinfo{person}{Lioba Heimbach}, {and} \bibinfo{person}{Roger
  Wattenhofer}.} \bibinfo{year}{2022}\natexlab{}.
\newblock \showarticletitle{{An Empirical Study of Market Inefficiencies in
  Uniswap and SushiSwap}}. In \bibinfo{booktitle}{\emph{{The 2nd Workshop on
  Decentralized Finance (DeFi), Grenada}}}.
\newblock


\bibitem[Black and Scholes(1973)]%
        {BlackScholes}
\bibfield{author}{\bibinfo{person}{Fischer Black} {and} \bibinfo{person}{Myron
  Scholes}.} \bibinfo{year}{1973}\natexlab{}.
\newblock \showarticletitle{The Pricing of Options and Corporate Liabilities}.
\newblock \bibinfo{journal}{\emph{The Pricing of Options and Corporate
  Liabilities}} \bibinfo{volume}{81}, \bibinfo{number}{3}
  (\bibinfo{year}{1973}), \bibinfo{pages}{637--654}.
\newblock


\bibitem[Chitra et~al\mbox{.}(2021)]%
        {chitra2021liveness}
\bibfield{author}{\bibinfo{person}{Tarun Chitra}, \bibinfo{person}{Guillermo
  Angeris}, {and} \bibinfo{person}{Alex Evans}.}
  \bibinfo{year}{2021}\natexlab{}.
\newblock \bibinfo{title}{How Liveness Separates CFMMs and Order Books}.
\newblock
\newblock


\bibitem[Corbet et~al\mbox{.}(2022)]%
        {corbet2022cryptocurrency}
\bibfield{author}{\bibinfo{person}{Shaen Corbet}, \bibinfo{person}{Yang~Greg
  Hou}, \bibinfo{person}{Yang Hu}, \bibinfo{person}{Charles Larkin},
  \bibinfo{person}{Brian Lucey}, {and} \bibinfo{person}{Les Oxley}.}
  \bibinfo{year}{2022}\natexlab{}.
\newblock \showarticletitle{Cryptocurrency liquidity and volatility
  interrelationships during the COVID-19 pandemic}.
\newblock \bibinfo{journal}{\emph{Finance Research Letters}}
  \bibinfo{volume}{45} (\bibinfo{year}{2022}), \bibinfo{pages}{102137}.
\newblock


\bibitem[Egorov(2019)]%
        {egorov2019stableswap}
\bibfield{author}{\bibinfo{person}{Michael Egorov}.}
  \bibinfo{year}{2019}\natexlab{}.
\newblock \bibinfo{title}{StableSwap-efficient mechanism for Stablecoin
  liquidity}.
\newblock
\newblock


\bibitem[Fritsch(2021)]%
        {Fritsch2021concentrated}
\bibfield{author}{\bibinfo{person}{Robin Fritsch}.}
  \bibinfo{year}{2021}\natexlab{}.
\newblock \showarticletitle{{Concentrated Liquidity in Automated Market
  Makers}}. In \bibinfo{booktitle}{\emph{{Proceedings of the 2021 ACM CCS
  Workshop on Decentralized Finance and Security (DeFi@CCS), Virtual Event,
  Republic of Korea}}}.
\newblock


\bibitem[Griffin and Shams(2020)]%
        {griffin2020bitcoin}
\bibfield{author}{\bibinfo{person}{John~M Griffin} {and} \bibinfo{person}{Amin
  Shams}.} \bibinfo{year}{2020}\natexlab{}.
\newblock \showarticletitle{Is Bitcoin really untethered?}
\newblock \bibinfo{journal}{\emph{The Journal of Finance}}
  \bibinfo{volume}{75}, \bibinfo{number}{4} (\bibinfo{year}{2020}),
  \bibinfo{pages}{1913--1964}.
\newblock


\bibitem[Heimbach et~al\mbox{.}(2022)]%
        {Heimbach2022risks}
\bibfield{author}{\bibinfo{person}{Lioba Heimbach}, \bibinfo{person}{Eric
  Schertenleib}, {and} \bibinfo{person}{Roger Wattenhofer}.}
  \bibinfo{year}{2022}\natexlab{}.
\newblock \showarticletitle{{Risks and Returns of Uniswap V3 Liquidity
  Providers}}. In \bibinfo{booktitle}{\emph{{4th ACM Conference on Advances in
  Financial Technologies (AFT), Cambridge, Massachusetts, USA}}}.
\newblock


\bibitem[Kliber(2018)]%
        {kliber2018price}
\bibfield{author}{\bibinfo{person}{Agata Kliber}.}
  \bibinfo{year}{2018}\natexlab{}.
\newblock \showarticletitle{Price, liquidity and information spillover within
  the cryptocurrency market. The case of bitfinex}.
\newblock \bibinfo{journal}{\emph{Safe Bank}} \bibinfo{volume}{73},
  \bibinfo{number}{4} (\bibinfo{year}{2018}), \bibinfo{pages}{62--79}.
\newblock


\bibitem[Loesch et~al\mbox{.}(2021)]%
        {loesch2021impermanent}
\bibfield{author}{\bibinfo{person}{Stefan Loesch}, \bibinfo{person}{Nate
  Hindman}, \bibinfo{person}{Mark~B Richardson}, {and}
  \bibinfo{person}{Nicholas Welch}.} \bibinfo{year}{2021}\natexlab{}.
\newblock \bibinfo{title}{Impermanent Loss in Uniswap v3}.
\newblock
\newblock


\bibitem[Neuder et~al\mbox{.}(2021)]%
        {neuder2021strategic}
\bibfield{author}{\bibinfo{person}{Michael Neuder}, \bibinfo{person}{Rithvik
  Rao}, \bibinfo{person}{Daniel~J Moroz}, {and} \bibinfo{person}{David~C
  Parkes}.} \bibinfo{year}{2021}\natexlab{}.
\newblock \bibinfo{title}{Strategic Liquidity Provision in Uniswap v3}.
\newblock
\newblock


\bibitem[Tang and Wang(2022)]%
        {tang2022liquidity}
\bibfield{author}{\bibinfo{person}{Tao Tang} {and} \bibinfo{person}{Yanchen
  Wang}.} \bibinfo{year}{2022}\natexlab{}.
\newblock \showarticletitle{Liquidity Shocks, Price Volatilities, and
  Risk-managed Strategy: Evidence from Bitcoin and Beyond}.
\newblock \bibinfo{journal}{\emph{Journal of Multinational Financial
  Management}} (\bibinfo{year}{2022}), \bibinfo{pages}{100729}.
\newblock


\bibitem[Wood(2014)]%
        {wood2014ethereum}
\bibfield{author}{\bibinfo{person}{Gavin Wood}.}
  \bibinfo{year}{2014}\natexlab{}.
\newblock \bibinfo{title}{Ethereum: A secure decentralised generalised
  transaction ledger}.
\newblock
\newblock


\end{thebibliography}
